\pdfoutput=1

\documentclass[11pt,twoside,a4paper,cmspaper,final,collab]{cms-tdr}
\def\svnVersion{445002P}\def\cmsCernNoTag{CERN-EP-2017-127}\def\cmsCernDate{\today}\def\cmsMessage{Published in Physical Review D as \href{http://dx.doi.org/10.1103/PhysRevD.97.032007}{\doi{10.1103/PhysRevD.97.032007.}}}

\begin{document}\cmsNoteHeader{SUS-16-044}

\hyphenation{had-ron-i-za-tion}
\hyphenation{cal-or-i-me-ter}
\hyphenation{de-vices}
\RCS$Revision: 443020 $
\RCS$HeadURL: svn+ssh://svn.cern.ch/reps/tdr2/papers/SUS-16-044/trunk/SUS-16-044.tex $
\RCS$Id: SUS-16-044.tex 443020 2018-01-27 11:39:29Z manuelf $

\newcommand{\cls}{\ensuremath{\mathrm{CL}_{\mathrm{S}}}\xspace}
\newcommand{\kapa}{\ensuremath{\kappa}\xspace}
\newcommand{\drmax}{\ensuremath{\Delta R_\text{max}}\xspace}
\newcommand{\amjj}{\ensuremath{\langle m\rangle}\xspace}
\newcommand{\dmjj}{\ensuremath{\Delta m}\xspace}
\newcommand{\njets}{\ensuremath{N_{\text{jets}}}\xspace}
\newcommand{\nb}{\ensuremath{N_{\PQb}}\xspace}
\newcommand{\nbl}{\ensuremath{N_{\PQb,\mathrm{L}}}\xspace}
\newcommand{\nbm}{\ensuremath{N_{\PQb,\mathrm{M}}}\xspace}
\newcommand{\nbt}{\ensuremath{N_{\PQb,\mathrm{T}}}\xspace}
\newcommand{\dphi}{\ensuremath{\Delta\phi}\xspace}
\newcommand{\mt}{\ensuremath{m_{\mathrm{T}}}\xspace}
\newcommand{\ptll}{\ensuremath{p_{\mathrm{T}}(\ell^{+}\ell^{-})}\xspace}
\newcommand{\mll}{\ensuremath{m(\ell^{+}\ell^{-})}\xspace}
\newcommand{\mLSP}{\ensuremath{m_{\PSGczDo}}\xspace}
\newcommand{\mGold}{\ensuremath{m_{\PXXSG}}\xspace}
\newcommand{\znn}{{\ensuremath{\cPZ\to\PGn\PAGn}}\xspace}
\newcommand{\zll}{{\ensuremath{\cPZ\to\ell^{+}\ell^{-}}}\xspace}
\newcommand{\zjets}{{\cPZ{}+jets}\xspace}
\newcommand{\wjets}{{\PW{}+jets}\xspace}
\newcommand{\ttjets}{{\ttbar{}+jets}\xspace}

\ifthenelse{\boolean{cms@external}}{\providecommand{\NA}{\ensuremath{\cdots}}}{\providecommand{\NA}{\ensuremath{\text{---}}}}
\ifthenelse{\boolean{cms@external}}{\providecommand{\CL}{C.L.\xspace}}{\providecommand{\CL}{CL\xspace}}
\newlength\cmsFigWidth
\ifthenelse{\boolean{cms@external}}{\setlength\cmsFigWidth{0.98\columnwidth}}{\setlength\cmsFigWidth{0.85\textwidth}}
\newlength\cmsFigFey
\ifthenelse{\boolean{cms@external}}{\setlength\cmsFigFey{0.8\linewidth}}{\setlength\cmsFigFey{0.42\textwidth}}
\newlength\cmsFigPanel
\ifthenelse{\boolean{cms@external}}{\setlength\cmsFigPanel{0.4\textwidth}}{\setlength\cmsFigPanel{0.49\textwidth}}
\newlength\cmsFigKappa
\ifthenelse{\boolean{cms@external}}{\setlength\cmsFigKappa{0.8\textwidth}}{\setlength\cmsFigKappa{\textwidth}}
\newlength\cmsFigLimit
\ifthenelse{\boolean{cms@external}}{\setlength\cmsFigLimit{0.7\textwidth}}{\setlength\cmsFigLimit{0.85\textwidth}}

\ifthenelse{\boolean{cms@external}}{\providecommand{\cmsLeft}{upper\xspace}}{\providecommand{\cmsLeft}{left\xspace}}
\ifthenelse{\boolean{cms@external}}{\providecommand{\cmsRight}{lower\xspace}}{\providecommand{\cmsRight}{right\xspace}}
\ifthenelse{\boolean{cms@external}}{\providecommand{\cmsTable[1]}{#1}}{\providecommand{\cmsTable}[1]{\resizebox{\textwidth}{!}{#1}}}
\cmsNoteHeader{SUS-16-044}
\title{\texorpdfstring{Search for higgsino pair production in $\Pp\Pp$ collisions at $\sqrt{s} = 13\TeV$ in final states with large missing transverse momentum and two Higgs bosons decaying via $\PH\to\bbbar$}{Search for higgsino pair production in pp collisions at sqrt(s) = 13 TeV in final states with large missing transverse momentum and two Higgs bosons decaying via H to bbbar}}

\date{\today}

\abstract{Results are reported from a search for new physics in 13\TeV proton-proton collisions in the final state with large missing transverse momentum and two Higgs bosons decaying via $\PH\to\bbbar$. The search uses a data sample accumulated by the CMS experiment at the LHC in 2016, corresponding to an integrated luminosity of 35.9\fbinv. The search is motivated by models based on gauge-mediated supersymmetry breaking, which predict the electroweak production of a pair of higgsinos, each of which can decay via a cascade process to a Higgs boson and an undetected lightest supersymmetric particle. The observed event yields in the signal regions are consistent with the standard model background expectation obtained from control regions in data. Higgsinos in the mass range 230--770\GeV are excluded at 95\% confidence level in the context of a simplified model for the production and decay of approximately degenerate higgsinos.}
\hypersetup{%
pdfauthor={CMS Collaboration},%
pdftitle={Search for higgsino pair production in pp collisions at sqrt(s) = 13 TeV in final states with large missing transverse momentum and two Higgs bosons decaying via H to bbbar},%
pdfsubject={CMS},%
pdfkeywords={CMS, physics, SUSY, Higgs}}

\maketitle

\section{Introduction}
\label{sec:intro}
The discovery of a Higgs boson~\cite{Aad:2012tfa,Chatrchyan:2012ufa,Chatrchyan:2013lba} at the electroweak
scale, with a mass $m_{\PH}\approx 125$\GeV ~\cite{Khachatryan:2014jba,Aad:2014aba,Aad:2015zhl}, provides a
new tool that can be used in searches for particles associated with physics beyond the standard model (SM).
Particles predicted by models based on supersymmetry
(SUSY)~\cite{Ramond:1971gb,Golfand:1971iw,Neveu:1971rx,Volkov:1972jx,Wess:1973kz,Wess:1974tw,Fayet:1974pd,Nilles:1983ge}
are expected in many cases to decay into Higgs bosons with significant branching fractions, and in some cases,
the presence of a Higgs boson can become a critical part of the experimental
signature~\cite{Khachatryan:2014mma,Khachatryan:2014qwa,Chatrchyan:2013mya,Khachatryan:2014doa,Aad:2015jqa}.

We perform a search for processes leading to Higgs boson pair production in association with large missing
transverse momentum, $\ptmiss$. Each Higgs boson is reconstructed in its dominant decay mode, $\PH\to\bbbar$,
which has a branching fraction of approximately 60\%. Such a signature can arise, for example, in models based
on SUSY, in which an electroweak process can lead to the production of two SUSY particles, each of which
decays into a Higgs boson and another particle that interacts so weakly that it escapes detection in the
apparatus. In this paper, we denote the particle in the search  signature as \PH because it corresponds to the
particle observed by ATLAS and CMS. However, in the context of  SUSY models such as the minimal supersymmetric
standard model (MSSM), this particle is usually assumed to correspond to the lighter of the two CP-even Higgs
particles, denoted as \Ph. The search uses an event sample of proton-proton ($\Pp\Pp$) collision data at $\sqrt
s=13$\TeV, corresponding to an integrated luminosity of 35.9\fbinv, collected by the CMS experiment at the
CERN LHC. Searches for this and related decay scenarios have been performed by
ATLAS~\cite{Aad:2012jva,Aad:2015jqa} and CMS~\cite{Khachatryan:2014qwa,Chatrchyan:2013mya,Khachatryan:2014mma}
using 7 and 8\TeV data.  The analysis described here is based on an approach developed in
Ref.~\cite{Khachatryan:2014mma}.

While the observation of a Higgs boson completes the expected spectrum of SM particles, the low value of its mass raises fundamental questions that suggest
the existence of new physics beyond the SM.  Assuming that the Higgs boson is a fundamental (that is,
noncomposite) spin-0 particle, stabilizing the electroweak scale (and the Higgs boson mass with it) is a major
theoretical challenge, referred to as the gauge hierarchy
problem~\cite{tHooft:1979bh,Witten:1981nf,Dine:1981za,Dimopoulos:1981au,Dimopoulos:1981zb,Kaul:1981hi}.
Without invoking new physics, the Higgs boson mass would be pulled by quantum loop corrections to the cutoff
scale of the theory, which can be as high as the Planck scale. Preventing such behavior requires an extreme
degree of fine tuning of the theoretical parameters. Alternatively, stabilization of the Higgs boson mass can
be achieved through a variety of mechanisms that introduce new physics at the \TeV scale,
such as SUSY.

The class of so-called natural SUSY
models~\cite{1988NuPhB.306...63B,Dimopoulos:1995mi,Barbieri:2009ev,Papucci:2011wy,Feng:2013pwa} contains the
ingredients necessary to stabilize the electroweak scale. These models are the object of an intensive program
of searches at the LHC. In any SUSY model, additional particles are introduced, such that all fermionic
(bosonic) degrees of freedom in the SM are paired with corresponding bosonic (fermionic) degrees of freedom in
the extended theory. In natural SUSY models, certain classes of partner particles are expected to be light.
These include the four higgsinos ($\widetilde{\mathrm{H}}^0_{1,2}$, $\widetilde{\mathrm{H}}^{\pm}$); both top squarks,
$\PSQt_L$ and $\PSQt_R$, which have the same electroweak couplings as the left- ($L$) and right- ($R$) handed
top quarks, respectively; the bottom squark with left-handed couplings ($\PSQb_L$); and the gluino ($\sGlu$).
Of these, the higgsinos are generically expected to be the lightest. Furthermore, in natural scenarios, the
four higgsinos are approximately degenerate in mass, so that transitions among these SUSY partners would
typically produce only very soft (\ie, low momentum) additional particles, which would not contribute to a
distinctive experimental signature.

In general, the gaugino and higgsino fields can mix, leading to mass eigenstates that are classified either as
neutralinos ($\widetilde{\chi}_i^0$, $i=$ 1--4) or charginos ($\widetilde{\chi}_i^{\pm}$, $i=$ 1--2). If the
\PSGczDo is the lightest SUSY particle (LSP), it is stable in models that conserve
$R$-parity~\cite{Farrar:1978xj} and, because of its weak interactions, would escape
experimental detection.  Achieving sensitivity to scenarios in which the higgsino sector is nearly mass
degenerate and contains the LSP poses a significant experimental challenge because the events are
characterized by low-\pt SM decay products and small values
of \ptmiss~\cite{Han:2014kaa,Han:2015lma,Baer:2014yta}.  Searches based on signatures involving initial-state
radiation (ISR) recoiling against the pair produced higgsinos have already excluded limited regions of phase
space for these
scenarios~\cite{Aad:2015zva,Khachatryan:2016mdm,Aaboud:2016tnv,Aaboud:2016qgg,Sirunyan:2017hci}. However,
achieving broad sensitivity based on this strategy is expected to require the large data samples that will be
accumulated by the HL-LHC~\cite{Han:2013usa}.

An alternative scenario arises, however, if the lightest higgsino/neutralino is not the LSP, but the
next-to-lightest SUSY particle (NLSP)~\cite{Ruderman:2011vv}. The LSP can be another particle that
is generic in SUSY models, the goldstino (\PXXSG). The goldstino is the Nambu--Goldstone particle
associated with the spontaneous breaking of global supersymmetry and is a fermion. In a broad range of models
in which SUSY breaking is mediated at a low scale, such as gauge-mediated supersymmetry breaking (GMSB)
models~\cite{Dimopoulos:1996vz,Matchev:1999ft}, the goldstino is nearly massless on the scale of the other
particles and becomes the LSP. If SUSY is promoted to a local symmetry, as is required for the full theory to
include gravity, the goldstino is ``eaten'' by the SUSY partner of the graviton, the gravitino ($J=3/2$), and
provides two of its four degrees of freedom.  In the region of parameter space involving prompt decays to the
gravitino, only the degrees of freedom associated with the goldstino have sufficiently large couplings to be
relevant, so it is common to denote the LSP in either case as a goldstino. In these GMSB models, the goldstino
mass is generically at the eV scale.

 \begin{figure}[tbp!]
\centering
\includegraphics[width=\cmsFigFey]{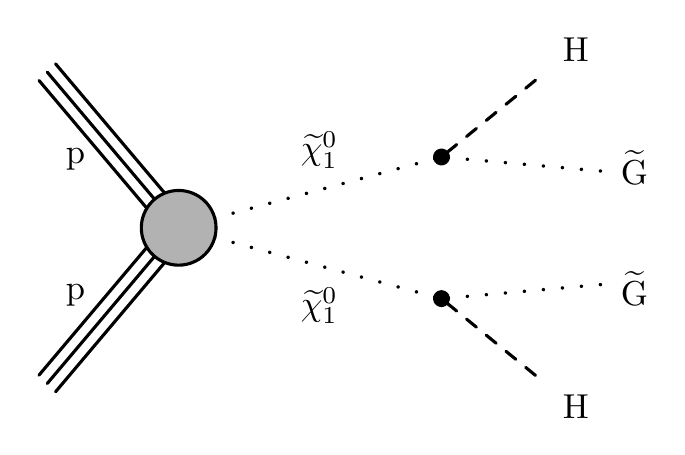}
\caption{Diagram for the gauge-mediated symmetry breaking signal model,
$\PSGczDo\PSGczDo\to\PH\PH\PXXSG\PXXSG$ (TChiHH), where \PXXSG is a goldstino.  The NLSPs \PSGczDo are not
directly pair produced, but are instead produced in the cascade decays of several different combinations of
neutralinos and charginos, as described in the text.}
\label{fig:intro:tchihh}
\end{figure}

If the lighter neutralinos and charginos are dominated by their higgsino content and are thus nearly mass
degenerate, their cascade decays can all lead to the production of the lightest neutralino,
\PSGczDo (now taken to be the NLSP), and soft particles.  Integrating over the contributions
from various allowed combinations of produced charginos and neutralinos
($\PSGczDo\PSGczDt$, $\PSGczDo\PSGcpmDo$,
$\PSGczDt\PSGcpmDo$, $\PSGcpmDo\widetilde{\chi}_1^{\mp}$) therefore leads to
an effective rate for $\PSGczDo\PSGczDo$ production~\cite{Fuks:2012qx,Fuks:2013vua} that
is significantly larger than that for any of the individual primary pairs, resulting in a boost to the
experimental sensitivity. The higgsino-like NLSP would then decay via $\PSGczDo\to \gamma\PXXSG$,
$\PSGczDo\to \PH\PXXSG$, or $\PSGczDo\to \PZ\PXXSG$, where the goldstino can lead to large
$\ptmiss$. The branching fractions for these decay modes vary depending on a number of parameters including
$\tan\beta$, the ratio of the Higgs vacuum expectation values, and the branching fraction for
$\PSGczDo\to\PH\PXXSG$ can be substantial. As a consequence, the signature ${\PH}{\PH}$+$\ptmiss$
with ${\PH}\to\bbbar$ can provide sensitivity to the existence of a higgsino sector in the important class of
scenarios in which the LSP mass lies below the higgsino masses.

This article presents a search for higgsinos in events with $\ptmiss>150$\GeV and at least three jets
identified as originating from \PQb quark hadronization (\PQb-tagged jets). In each event, we reconstruct two
Higgs boson candidates and define search regions within a Higgs boson mass window. The background is dominated
by \ttbar production at low and intermediate \ptmiss, and by \znn production in association with \PQb quarks
at high \ptmiss. The background is estimated entirely from data control regions corresponding to events with
two \PQb-tagged jets and events with three or four \PQb-tagged jets outside the Higgs boson mass
window. Systematic uncertainties on the background prediction are derived from both dedicated data control
regions for \ttbar, \znn and QCD multijet production as well as from the simulation of the background events
in the search regions.

Results are interpreted in the simplified model framework~\cite{bib-sms-2,bib-sms-3,bib-sms-4} using the model
shown in Fig.~\ref{fig:intro:tchihh}, hereafter referred to as TChiHH. In this model, two $\PSGczDo$
NLSPs are produced, each decaying via $\PSGczDo\to{\PH}\PXXSG$. The cross-section is calculated
under the assumption that at least one of the $\PSGczDo$ NLSPs is produced in a cascade decay of
$\PSGczDt$ or $\PSGcpmDo$, as described above.  This situation arises when the mass splittings
among charginos and neutralinos are large enough that the decays to $\PSGczDo$ occur
promptly~\cite{Meade:2009qv}, while also small enough that the additional soft particles fall out of
acceptance.

\section{The CMS detector}
\label{sec:detector}

The central feature of the CMS detector is a superconducting solenoid of 6\unit{m} internal diameter,
providing a magnetic field of 3.8\unit{T}. Within the solenoid volume are the tracking and
calorimeter systems. The tracking system, composed of silicon-pixel and silicon-strip
detectors, measures charged particle trajectories within
the pseudorapidity range $\abs{\eta}<2.5$, where $\eta\equiv -\ln[\tan(\theta/2)]$ and $\theta$ is the
polar angle of the trajectory of the particle with respect to the counterclockwise proton beam direction.
A lead tungstate crystal electromagnetic calorimeter (ECAL), and a brass and scintillator hadron calorimeter (HCAL),
each composed of a barrel and two endcap sections, provide energy measurements up to $\abs{\eta}=3$.
Forward calorimeters extend the pseudorapidity coverage provided by the barrel and
endcap detectors up to $\abs{\eta}=5$. Muons are identified and measured within the range $\abs{\eta}<2.4$
by gas-ionization detectors embedded in the steel flux-return yoke outside the solenoid.
The detector is nearly hermetic, permitting the accurate measurement of $\ptmiss$.
A more detailed description of the CMS detector, together with a definition of the coordinate
system used and the relevant kinematic variables, is given in Ref.~\cite{Chatrchyan:2008zzk}.

\section{Simulated event samples}
\label{sec:samples}
Several simulated event samples are used for modeling the SM background and signal processes.
While the background estimation in the analysis is performed from control samples in the data,
simulated event samples are used to estimate uncertainties, as well as to build an understanding of the
characteristics of the selected background events.

The production of \ttjets, \wjets, \zjets, and quantum chromodynamics (QCD) multijet events is simulated with
the Monte Carlo (MC) generator \MGvATNLO~2.2.2~\cite{Alwall:2014hca} in leading-order (LO)
mode~\cite{Alwall:2007fs}. Single top quark events are modeled with \POWHEG
v2~\cite{Alioli:2009je,Re:2010bp} for the $t$-channel and $t$\PW\ production,
and \MGvATNLO at
next-to-leading order (NLO)~\cite{Frederix:2012ps} for the $s$-channel.  Additional
small backgrounds, such as \ttbar production in association with bosons, diboson processes, and
$\ttbar\ttbar$, are also produced at NLO with either \MGvATNLO or \POWHEG. All events are
generated using the NNPDF~3.0~\cite{Ball:2014uwa} set of parton distribution functions. Parton showering and
fragmentation are performed with the \PYTHIA~8.205~\cite{Sjostrand:2014zea} generator with the underlying
event model based on the CUETP8M1 tune~\cite{Khachatryan:2110213}. The detector simulation is performed with
\GEANTfour~\cite{Agostinelli:2002hh,Allison:2006ve,Allison:2016lfl}. The cross sections used to scale
simulated event yields are based on the highest order calculation available~\cite{
Beneke:2011mq,        
Cacciari:2011hy,      
Baernreuther:2012ws,  
Czakon:2012zr,        
Czakon:2012pz,        
Czakon:2013goa,       
Gavin:2012sy,         
Gavin:2010az,         
Alioli:2009je,  
Re:2010bp,      
Melia:2011tj},    
which for the most part correspond to NLO or next-to-NLO precision.

Signal events for the TChiHH simplified model are generated for 36 values of the higgsino mass between 127 and
1000\GeV. The mass points are denoted as TChiHH(\mLSP,\mGold), where \mLSP is the higgsino mass and \mGold is
the mass of the LSP, both in units of \GeV. While the value of \mGold is fixed to 1\GeV in the simulation for
technical reasons, the resulting event kinematics are consistent with an approximately massless LSP such as
the goldstino in GMSB.  The yields are normalized to the NLO + next-to-leading logarithmic (NLL) cross
section~\cite{Fuks:2012qx,Fuks:2013vua}.  The production cross sections are calculated in the limit of mass
degeneracy among higgsinos, $\PSGczDo$, $\PSGczDt$, and $\PSGcpmDo$. All other supersymmetric partners of the
SM particles are assumed to be heavy (100\TeV) and thus essentially decoupled, a simplification that has a
very small impact on higgsino pair production (e.g.,~the cross section changes less than 2\% when the masses of
the gluino and squarks are lowered down to 500\GeV).  Following the convention of real mixing matrices and
signed neutralino masses~\cite{Skands:2003cj}, we set the sign of the mass of $\PSGczDo$ ($\PSGczDt$) to $+1$
($-1$).  The lightest two neutralino states are defined as symmetric (anti-symmetric) combinations of higgsino
states by setting the product of the elements $N_{i3}$ and $N_{i4}$ of the neutralino mixing matrix $N$ to
$+0.5$ ($-0.5$) for $i = 1$ ($2$).  The elements $U_{12}$ and $V_{12}$ of the chargino mixing matrices $U$ and
$V$ are set to 1.  All chargino and neutralino decays in the simplified model are taken to be prompt, although
the lifetimes of particles in a physical model would depend on the mass splitting between the higgsino-like
states, which become long-lived in the limit of degenerate masses.  Both Higgs bosons in each event are forced
to decay to \bbbar, which is accounted for by scaling the signal event yields with the branching fraction of
58.24\%~\cite{deFlorian:2016spz}. The signal contribution from Higgs boson decays other than $\PH\to\bbbar$ is
small in this analysis and is ignored. Signal events are generated in a manner similar to that for the SM
backgrounds, with the \MGvATNLO~2.2.2 generator in LO mode using the NNPDF~3.0 set of parton distribution
functions and followed by \PYTHIA~8.205 for showering and fragmentation.  The detector simulation is performed
with the CMS fast simulation package~\cite{Abdullin:2011zz} with scale factors applied to compensate for any
differences with respect to the full simulation.

Finally, to model the presence of additional $\Pp\Pp$ collisions from the same beam
crossing as the primary hard-scattering process or another beam crossing adjacent to it
(``pileup'' interactions), the simulated events are overlaid with
multiple minimum-bias events, which are also generated with the \PYTHIA~8.205 generator with the underlying
event model based on the CUETP8M1 tune.

\section{Event reconstruction}
\label{sec:objects}

\subsection{Object definitions}

The reconstruction of physics objects in an event proceeds from the candidate particles identified by the
particle-flow (PF) algorithm~\cite{Sirunyan:2017ulk}, which uses information from the
tracker, calorimeters, and muon systems to identify the candidates as charged or neutral hadrons, photons,
electrons, or muons.  The reconstructed vertex with the largest value of summed physics-object $\pt^2$, with
\pt denoting transverse momentum, is taken to be the primary $\Pp\Pp$ interaction vertex (PV).  The physics
objects used in this context are the objects returned by a jet finding
algorithm~\cite{Cacciari:2008gp,Cacciari:2011ma} applied to all charged tracks associated with the
vertex under consideration, plus the corresponding associated \ptmiss.

The charged PF candidates associated with the PV and the neutral PF candidates are clustered into jets using
the anti-$\kt$ algorithm~\cite{Cacciari:2008gp} with a distance parameter $R=0.4$, as implemented in the
\FASTJET package~\cite{Cacciari:2011ma}. The jet momentum is determined as the vectorial sum of all
particle momenta in the jet. Jet energy corrections are derived based on a combination of simulation studies,
accounting for the nonlinear detector response and the presence of pileup, together with in situ measurements
of the \pt balance in dijet and $\gamma$+jet events~\cite{Khachatryan:2016kdb}. The resulting calibrated
jet is required to satisfy $\pt>30\GeV$ and $\abs{\eta}\leq2.4$.  Additional selection criteria are applied to
each event to remove spurious jet-like features originating from isolated noise in certain HCAL
regions~\cite{CMS-PAS-JME-10-003}.

A subset of the jets are ``tagged'' as originating from \PQb quarks using \textsc{DeepCSV}~\cite{Sirunyan:2017ezt}, a
new \PQb tagging algorithm based on a deep neural network with four hidden layers~\cite{Guest:2016iqz}.
We use all three of the
\textsc{DeepCSV} algorithm working points: loose, medium, and tight, defined by the values of the
discriminator requirement for which the rates for misidentifying a light-flavor jet as a \PQb jet are 10\%,
1\%, and 0.1\%, respectively. The \PQb tagging efficiency for jets with $\pt$ in the 80--150\GeV range is
approximately 86\%, 69\%, and 51\% for the loose, medium, and tight working points, respectively, and
gradually decreases for lower and higher jet \pt. The simulation is reweighted to
compensate for any differences with respect to data based on measurements of the \PQb tagging efficiency and
mistag rate for each working point in dedicated data samples~\cite{Sirunyan:2017ezt}.

The missing transverse momentum, \ptmiss, is given by the magnitude of $\ptvecmiss$, the negative vector sum
of the transverse momenta of all PF candidates~\cite{cms-pas-pft-09-001,cms-pas-pft-10-001}, adjusted for
known detector effects by taking into account the jet energy corrections. Filters are applied to reject events
with well defined anomalous sources of \ptmiss arising from calorimeter noise, dead calorimeter cells, beam halo, and
other effects.

Since the targeted signature is fully hadronic, contamination from final states involving leptons in the
search region is suppressed by vetoing events with reconstructed lepton candidates. Electrons are identified
by associating a charged particle track with an ECAL supercluster~\cite{Khachatryan:2015hwa} and are required
to have $\pt>10\GeV$ and $\abs{\eta}<2.5$.  Muons are identified by associating tracks in the muon system with
those found in the silicon tracker~\cite{Chatrchyan:2012xi} and are required to satisfy $\pt>10\GeV$ and
$\abs{\eta}<2.4$.

To preferentially consider only leptons that originate in the decay of $\PW$ and $\cPZ$ bosons, leptons are
required to be isolated from other PF candidates using an optimized version of the
``mini-isolation''~\cite{Rehermann:2010vq,Khachatryan:2016uwr}. The isolation $I_\text{mini}$ is obtained by
summing the transverse momentum of the charged hadrons, neutral hadrons, and photons within $\Delta
R\equiv\sqrt{\smash[b]{(\Delta\phi)^2+(\Delta\eta)^2}}<R_0$ of the lepton momentum vector $\vec{p}^{\ell}$, where
$\phi$ is the azimuthal angle in radians and $R_0$ is given by 0.2 for $p^{\ell}_\mathrm{T}\leq 50\GeV$,
$(10\GeV)/p^{\ell}_\mathrm{T}$ for $50 <p^{\ell}_\mathrm{T}< 200\GeV$, and 0.05 for $p^{\ell}_\mathrm{T}\geq
200\GeV$. Electrons (muons) are then required to satisfy $I_\text{mini}/p^{\ell}_\mathrm{T}< 0.2\,(0.1)$.

As described in Section~\ref{sec:eventSelection}, the
dominant background arises from the production of single-lepton \ttbar events in which the lepton is a $\tau$
decaying hadronically, or is a light lepton that is either not reconstructed or fails the lepton selection
criteria, including the \pt threshold and the isolation requirements. To reduce this background, we veto
events with any additional isolated tracks corresponding to leptonic or hadronic PF candidates.  To reduce the
influence of tracks originating from pileup, isolated tracks are considered only if their closest distance of
approach along the beam axis to a reconstructed vertex is smaller for the primary event vertex than for any
other vertex.

The requirements for the definition of an isolated track differ slightly depending on whether the track is
identified as leptonic or hadronic by the PF algorithm.  For leptonic tracks, we require
$\pt>5\gev$ and $I_\text{trk}<0.2$,
where $I_\text{trk}$ is the scalar \pt sum of other charged tracks within $\Delta
R<0.3$ of the primary track, divided by the \pt value of the primary
track.  For hadronic tracks, we apply slightly tighter requirements to reduce hadronic (non-$\tau$) signal
loss: $\pt>10\gev$ and $I_\text{trk}<0.1$.
To minimize the signal inefficiency due to this veto, isolated tracks are
considered only if they are consistent with originating from a \PW\ boson decay, specifically, if the transverse mass of the track and the missing transverse momentum satisfy
\ifthenelse{\boolean{cms@external}}{
\begin{linenomath*}
\begin{multline}
\label{eq:mt_isotk}
m_\mathrm{T}(\vec{p}^{\, \text{trk}}_\mathrm{T},\ptvecmiss)\\
 \equiv \sqrt{2p_\mathrm{T}^\text{trk}\ptmiss[1-\cos(\Delta\phi_{\vec{p}^{\,\text{trk}}_\mathrm{T}, \ptvecmiss} )]}
<100\GeV,
\end{multline}
\end{linenomath*}
}{
\begin{linenomath*}
\begin{equation}
\label{eq:mt_isotk}
m_\mathrm{T}(\vec{p}^{\, \text{trk}}_\mathrm{T},\ptvecmiss) \equiv \sqrt{2p_\mathrm{T}^\text{trk}\ptmiss[1-\cos(\Delta\phi_{\vec{p}^{\,\text{trk}}_\mathrm{T}, \ptvecmiss} )]}
<100\GeV,
\end{equation}
\end{linenomath*}
}
where $\vec{p}^{\,\text{trk}}_\mathrm{T}$ is the transverse momentum of the track and $\Delta\phi_{\vec{p}^{\,\text{trk}}_\mathrm{T}, \ptvecmiss}$ is
the azimuthal separation between the track and \ptvecmiss.

The majority of QCD multijet events containing high \ptmiss have at least one jet with undermeasured
momentum and thus a spurious momentum imbalance.  A signature of such an event is a jet closely aligned in
direction with the \ptvecmiss vector.  To suppress this background, we place the following requirements on the
angle $\Delta\phi_i$ between the $i$th highest \pt jet and \ptvecmiss for $i=1$, 2, 3, 4: $\Delta\phi_1 > 0.5$,
$\Delta\phi_2 > 0.5$,  $\Delta\phi_3 > 0.3$, and $\Delta\phi_4 > 0.3$. These conditions are hereafter
collectively referred to as the high $\Delta\phi$ requirement.

The number of jets satisfying the criteria described above is denoted as \njets, while the numbers of these
jets tagged with the loose, medium, and tight \PQb tagging working points are labeled as \nbl, \nbm, and
\nbt, respectively. By definition, the jets identified by each \PQb tagging working point form a subset of
those satisfying the requirements of looser working points.

\subsection{Definition of \texorpdfstring{\PQb}{b} tag categories}

To optimize signal efficiency and background rejection, we define the following mutually exclusive
\PQb tag categories:
\begin{itemize}
  \item 2\PQb category: $\nbt=2$, $\nbm=2$, $\nbl\geq2$,
  \item 3\PQb category: $\nbt\geq2$, $\nbm=3$, $\nbl=3$, and
  \item  4\PQb  category: $\nbt\geq2$, $\nbm\geq3$, $\nbl\geq4$.
\end{itemize}

The 2\PQb category is used as a control sample to determine the kinematic shape of the background. Most of the
signal events lie in the 3$\PQb$ and  4\PQb  categories.
This categorization is found to have superior performance with respect to other combinations of
working points.  For instance, the simpler option of only using medium $\PQb$ tags results in a 2$\PQb$
control sample that has a larger contribution from QCD multijet production and a 4$\PQb$ sample with smaller signal efficiency.

To study various sources of background with higher statistical precision, we also define the following
\PQb tag categories with looser requirements:
\begin{itemize}
  \item 0\PQb category: $\nbm=0$,
  \item 1\PQb category: $\nbm=1$.
\end{itemize}
We will use \nb as a shorthand when discussing \PQb tag categories as an analysis variable, and \nbl, \nbm, and
\nbt when discussing numbers of \PQb-tagged jets for specific working points.

\subsection{Higgs boson pair reconstruction}
\label{sec:obj:higgs}

The principal visible decay products in signal events are the four \PQb jets from the decay of the
two Higgs bosons.  Additional jets may arise from initial- or final-state radiation as well as
pileup. To reconstruct both Higgs bosons, we choose the four jets with the largest \textsc{DeepCSV}
discriminator values,
i.e.,~the four most \PQb-quark-like jets.  These four jets can be grouped into three different pairs of Higgs boson
candidates.  Of the three possibilities, we choose the one with the smallest mass difference \dmjj between the
two Higgs boson candidate masses $m_{\PH_1}$ and $m_{\PH_2}$,
\begin{linenomath*}
\begin{equation}
\dmjj \equiv \left|  m_{\PH_1}-m_{\PH_2} \right|.
\end{equation}
\end{linenomath*}
This method exploits the fact that signal events contain two particles of identical mass, without using the
known value of the Higgs boson mass itself.  Methods that use the known mass to select the best candidate tend
to sculpt an artificial peak in the background.

Only events where the masses of the two Higgs boson candidates are similar, $\dmjj<40$\GeV, are kept. We
then calculate the average mass as
\begin{linenomath*}
\begin{equation}
\amjj \equiv \frac{m_{\PH_1}+m_{\PH_2}}{2}.
\end{equation}
\end{linenomath*}
As discussed in Section~\ref{sec:backgroundEstimation}, the search is then performed within the Higgs boson
mass window defined as $100<\amjj\leq140$\GeV.

After selecting the two Higgs boson candidates, we compute the distance $\Delta R$
between the two jets in each of the $\PH\to\bbbar$ candidate decays.
We then define \drmax as the larger of these two values,
\begin{linenomath*}
\begin{equation}
\drmax \equiv \max\left(\Delta R_{\PH_1}, \Delta R_{\PH_2} \right).
\end{equation}
\end{linenomath*}

In the typical configuration of signal events satisfying the baseline requirements, \drmax is small
because the Higgs bosons tend to have nonzero transverse boost and, thus, the two jets from the decay of a
Higgs boson tend to lie near each other in $\eta$ and $\phi$.  In contrast, for semileptonic \ttbar background
events, three of the jets typically arise from a top quark that decays via a hadronically decaying {\PW} boson
while the fourth jet arises from a \PQb quark from the other top quark decay.  Therefore, three of the jets
tend to lie within the same hemisphere while the fourth jet is in the opposite hemisphere.  One of the Higgs
boson candidates is thus formed from jets in both hemispheres, and \drmax tends to be larger than it is for
signal events.

\section{Trigger and event selection}
\label{sec:eventSelection}

The data sample was obtained with triggers that require the online \ptmiss value to be greater than 100 to
120\GeV, the applied threshold varying with the instantaneous luminosity delivered by the LHC. This variable
is computed with trigger-level quantities, and therefore has somewhat worse resolution than the corresponding
offline variable. The trigger efficiency measured as a function of offline \ptmiss, in samples triggered by a
high-\pt isolated electron, rises rapidly from about 60\% at $\ptmiss = 150$\GeV to 92\% for $\ptmiss =
200$\GeV and to over 99\% for $\ptmiss>300$\GeV. The systematic uncertainty in the trigger efficiency is
obtained by comparing the nominal efficiency with that found in different kinematic regions, with various
reference triggers, and with the simulation. This uncertainty is about 7\% for $\ptmiss=150$\GeV and decreases
to 0.7\% for $\ptmiss>300$\GeV.

Several data control samples are employed to validate the analysis techniques and to estimate systematic
uncertainties in the background estimates. The control sample for the principal background from \ttbar events
requires exactly one electron or one muon, while the \znn background is studied with a control sample
requiring two leptons consistent with a \zll decay.  These data samples were obtained with
triggers that require at least one electron or muon with \pt greater than 27 or 24\GeV, respectively.

\begin{figure*}[tbp!]
\centering
\includegraphics[width=\cmsFigPanel]{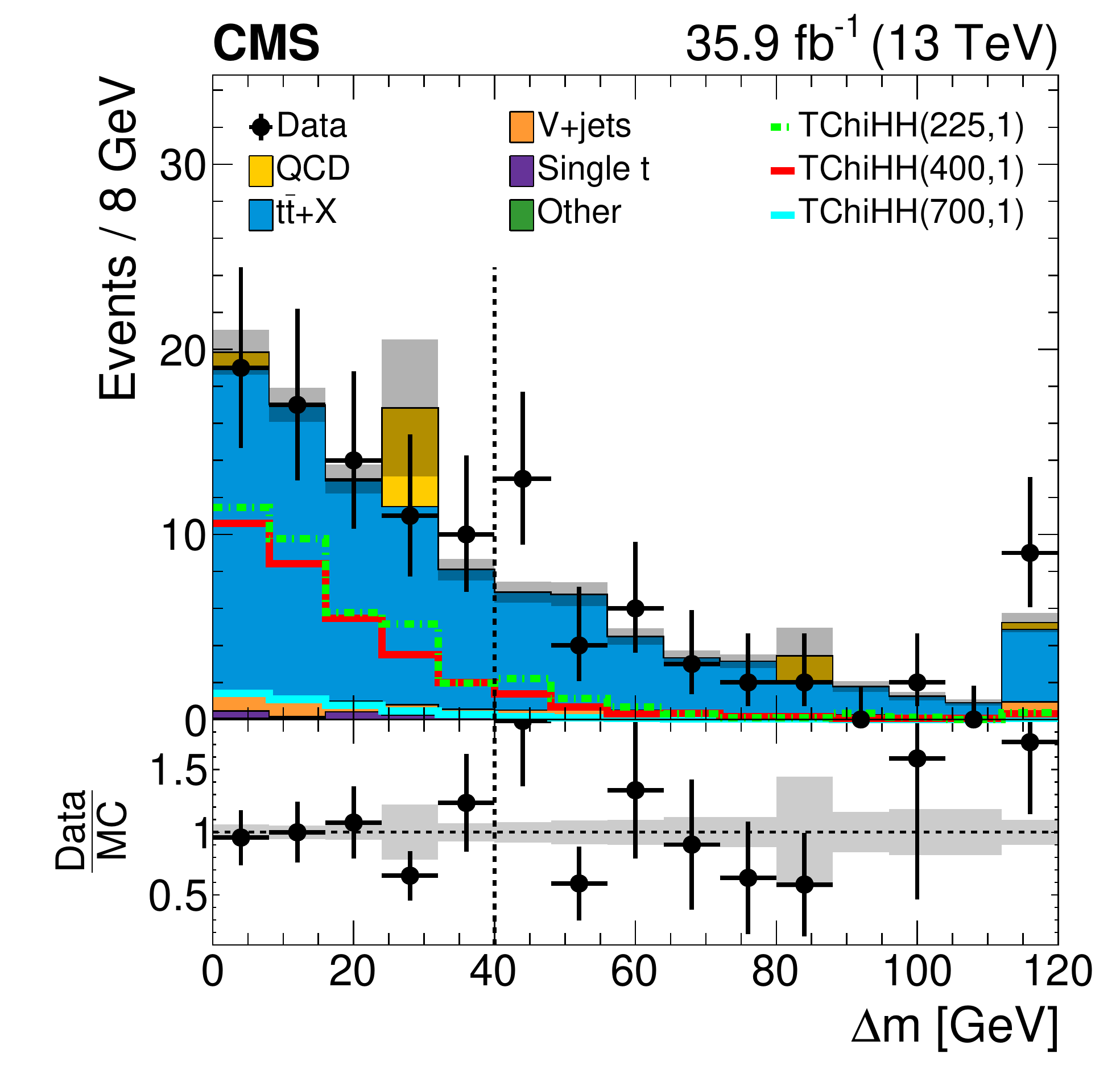}
\includegraphics[width=\cmsFigPanel]{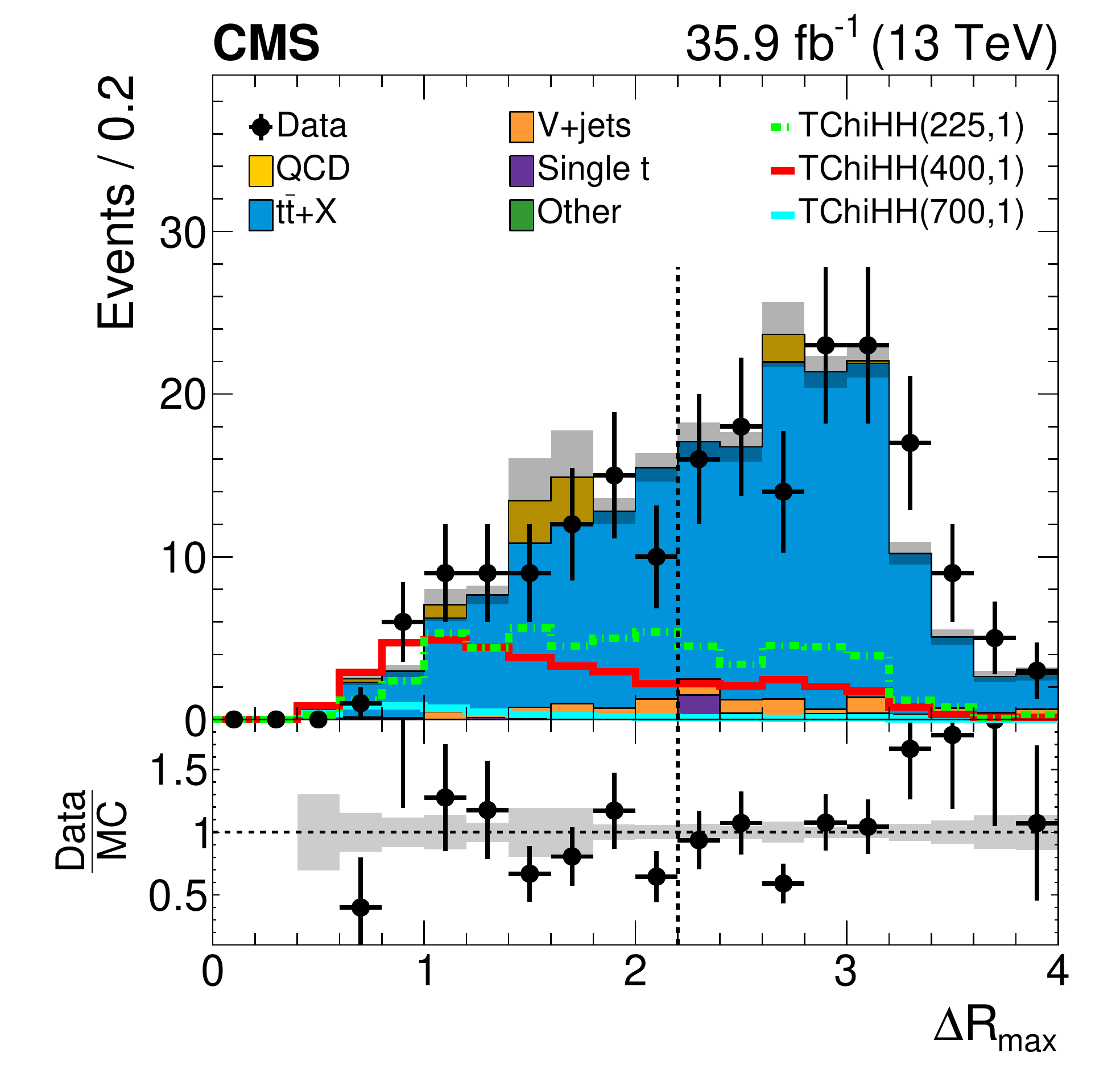}
\includegraphics[width=\cmsFigPanel]{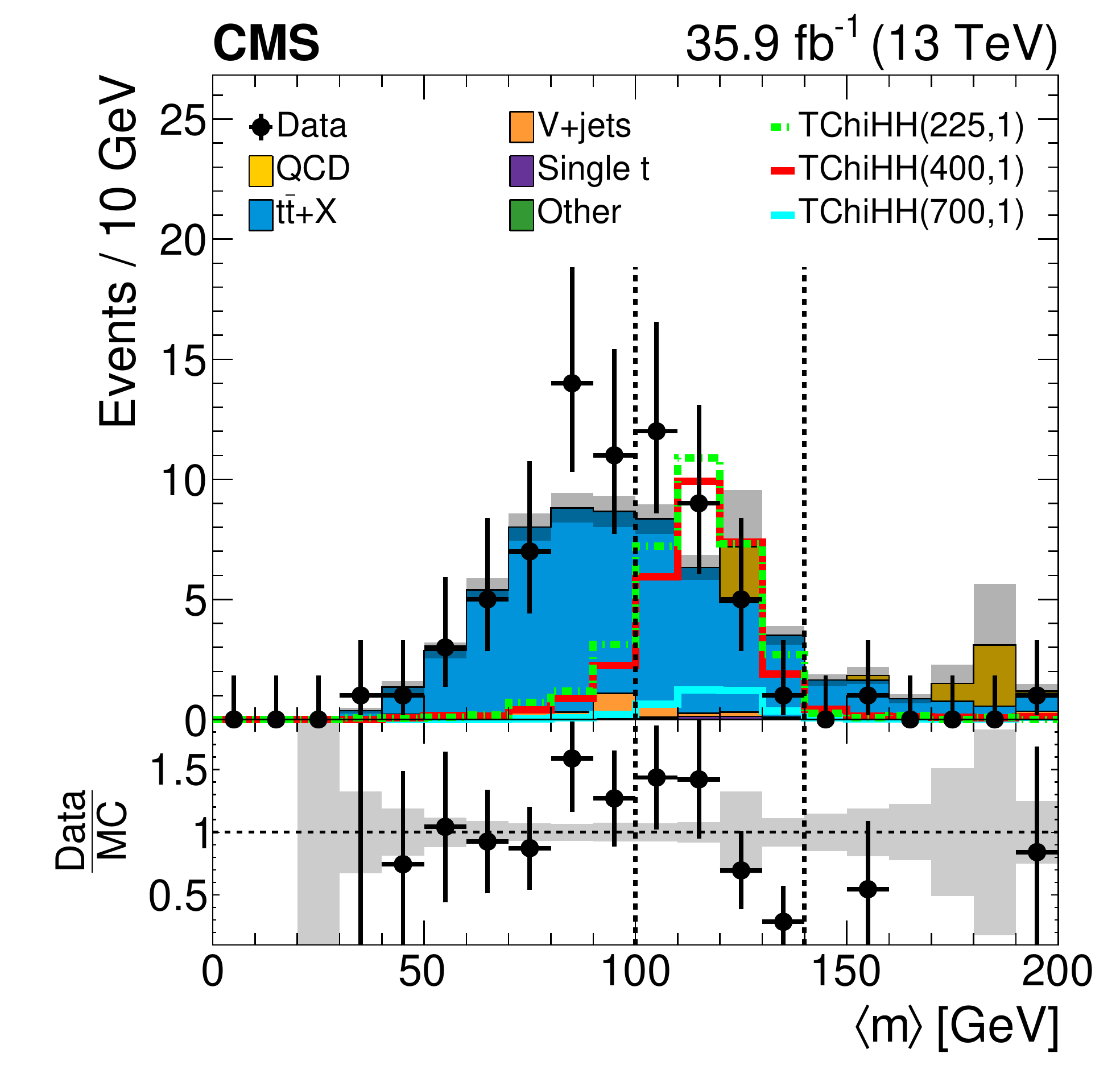}
\includegraphics[width=\cmsFigPanel]{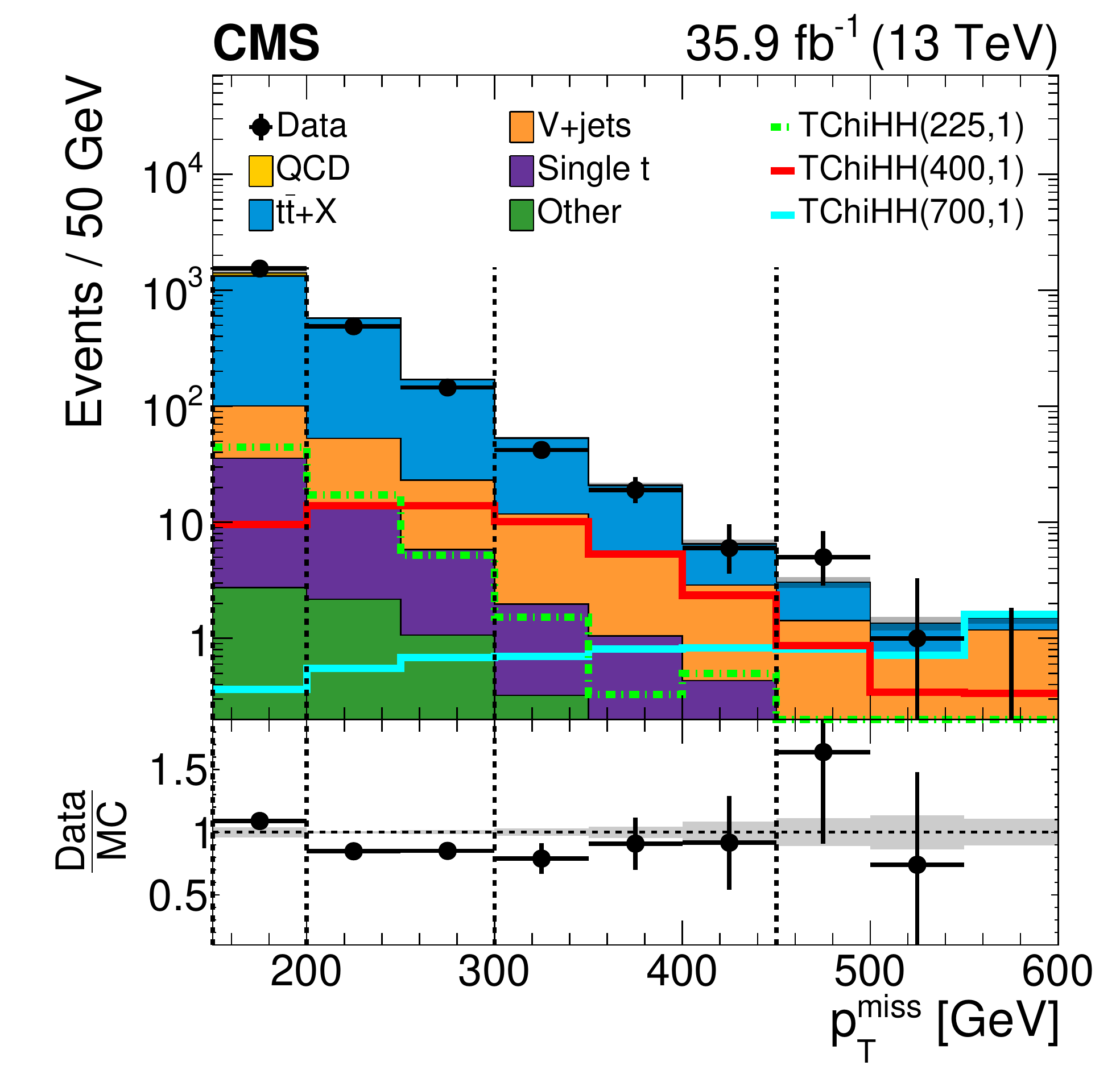}
\caption{Distributions of \dmjj, \drmax, \amjj, and \ptmiss for data  and simulated background samples,
  as well as three signal benchmark points denoted as TChiHH(\mLSP,\mGold), with \mLSP and \mGold in units of \GeV. All figures include baseline
    requirements (except on the variable being plotted in the case of \dmjj and \drmax). The \dmjj, \amjj,
    and \drmax distributions also include the  4\PQb  selection. The simulation is normalized to the observed data
    yields. The gray shading indicates the statistical uncertainty in the total simulated background. The
    vertical dotted lines indicate baseline requirements in the top row figures, the search region mass window
    in \amjj in the bottom left figure, and the \ptmiss binning in the bottom right figure. The last bin
    includes overflow.}
\label{fig:seln:nminus1}
\end{figure*}

Signal events have four \PQb jets from the decay of two Higgs bosons and no isolated leptons, with any
additional hadronic activity coming from initial- or final-state radiation. Thus, we select events with four
or five jets, no leptons or isolated tracks, $\nbt\geq2$, $\ptmiss>150$\GeV, high \dphi, $\dmjj<40$\GeV, and
$\drmax<2.2$. These selection requirements, listed in the top half of Table~\ref{tab:selection:cutflow}, are
referred to as the \textit{baseline selection}, while the bottom half of that table shows the further reduction
in background in increasingly more sensitive search bins. The distributions of \dmjj, \drmax, and
\amjj  in the  4\PQb  category are shown in Fig.~\ref{fig:seln:nminus1} in data and simulation. The actual
background prediction, however, is based on data control samples, as described in the next section.

\begin{table*}[tbh]
\centering
\topcaption{Event yields obtained from simulated event samples scaled to an integrated luminosity of 35.9\fbinv,
as the event selection criteria are applied.  The category ``$\ttbar$+$X$'' is dominated by \ttbar (98.5\%),
but also includes small contributions from $\ttbar\ttbar$, $\ttbar\PW$, $\ttbar\cPZ$, $\ttbar\PH$, and
$\ttbar\gamma$ backgrounds.  The category ``V+jets'' includes \zjets and \wjets backgrounds in all their decay
modes.  The category ``Other'' includes $\cPZ\cPZ$, $\PW\cPZ$, $\PW\PW$, $\PW\PH(\to \bbbar)$, and
$\cPZ\PH(\to \bbbar)$ processes.  The event selection requirements listed up to and including $\Delta
R_{\text{max}} < 2.2$ are defined as the {\it baseline selection}. The trigger efficiency is applied as an
event weight and is first taken into account in the $\ptmiss>150$\GeV row. The uncertainties in the ``Total SM
bkg.'' column is statistical only. The columns corresponding to the yields for three signal benchmark points
are labeled by TChiHH(\mLSP,\mGold), with \mLSP and \mGold in units of \GeV. The simulated samples for
TChiHH(225,1), TChiHH(400,1), and TChiHH(700,1) are equivalent to 10, 100, and over 1000 times the data
sample, respectively, so the statistical uncertainties in the signal yields are small.}
\label{tab:selection:cutflow}
\cmsTable{
\newcolumntype{x}{D{,}{.}{5.7}}
  \begin{scotch}{ l | rrrrr | x | r  r  r }
 & & & & & & & TChiHH & TChiHH & TChiHH\\
 \multicolumn{1}{c|}{$\mathcal{L} = 35.9$\fbinv}  & Other & Single t & QCD & V+jets & $\ttbar$+$X$ & \multicolumn{1}{c|}{Total SM bkg.} & (225,1) & (400,1) & (700,1)\\
\hline
    No selection                 & \multicolumn{1}{c}{\NA} & \multicolumn{1}{c}{\NA} & \multicolumn{1}{c}{\NA} & \multicolumn{1}{c}{\NA} & \multicolumn{1}{c|}{\NA} & \multicolumn{1}{c|}{\NA} & 10477.0 & 1080.3 & 84.0\\
    $0\ell$, $\text{4--5 jets}$   & \multicolumn{1}{c}{\NA} & \multicolumn{1}{c}{\NA} & \multicolumn{1}{c}{\NA} & \multicolumn{1}{c}{\NA} & \multicolumn{1}{c|}{\NA} & \multicolumn{1}{c|}{\NA} & 4442.0 & 544.9 & 44.6\\
    $\nbt\geq 2$       & \multicolumn{1}{c}{\NA} & \multicolumn{1}{c}{\NA} & \multicolumn{1}{c}{\NA} & \multicolumn{1}{c}{\NA} & \multicolumn{1}{c|}{\NA} & \multicolumn{1}{c|}{\NA} & 2509.3 & 308.9 & 23.9\\
    $\ptmiss>150$\GeV & 122.3 & 1847.0 & 13201.4 & 2375.8 & 26797.7 & 44344,2\pm778.5 & 509.5 & 204.2 & 20.4\\
    Track veto & 91.4 & 1130.1 & 12251.8 & 1987.0 & 16910.1 & 32370,5\pm770.5 & 476.9 & 196.3 & 19.9\\
    High $\Delta\phi$ & 62.3 & 688.4 & 1649.0 & 1466.6 & 12027.0 & 15893,4\pm482.6 & 267.2 & 162.3 & 17.5\\
    $|\Delta m| < 40$\GeV & 35.9 & 366.0 & 831.9 & 745.5 & 7682.3 & 9661,6\pm440.8 & 191.8 & 119.4 & 12.2\\
    $\Delta R_{\text{max}} < 2.2$ & 14.2 & 138.2 & 147.0 & 336.9 & 3014.2 & 3650,5\pm90.2 & 98.3 & 79.6 & 10.1\\[1.8ex]
    $100<\left< m \right>\leq140$\GeV & 3.8 & 42.3 & 14.0 & 75.2 & 992.0 & 1127,3\pm10.1 & 72.9 & 61.0 & 8.3\\
    3\PQb +  4\PQb  & 0.1 & 3.4 & 3.2 & 7.1 & 109.0 & 122,9\pm3.9 & 54.9 & 46.5 & 6.3\\
     4\PQb  & 0.1 & 0.7 & 3.2 & 1.5 & 27.3 & 32,8\pm3.4 & 38.1 & 32.8 & 4.6\\
    $\ptmiss>200$\GeV & 0.1 & 0.3 & 3.2 & 1.1 & 9.4 & 14,1\pm3.3 & 16.2 & 27.4 & 4.3\\
    $\ptmiss>300$\GeV & 0.0 & 0.1 & 0.0 & 0.4 & 1.1 & 1,7\pm0.2 & 2.0 & 11.5 & 3.5\\
    $\ptmiss>450$\GeV & 0.0 & 0.0 & 0.0 & 0.1 & 0.1 & 0,1\pm0.1 & 0.0 & 1.1 & 2.0\\
  \end{scotch}
}
\end{table*}

Based on the simulation, after the baseline selection, more than 85\% of the remaining SM background arises
from semileptonic \ttbar production. Approximately half of this contribution corresponds to \ttbar events with
an electron or a muon in the final state that is either out of acceptance or not identified, while the other
half involves final states with a hadronically decaying $\tau$ lepton. The contributions from events with
a \PW\ or \cPZ\ boson in association with jets (V$+$jets) are about 10\% and are dominated by \znn decays. The
background from QCD multijet events after the baseline selection is very small due to the combination
of \ptmiss, $\Delta\phi$, and \nb requirements.

As shown in Fig.~\ref{fig:seln:nminus1}, the \ptmiss distribution of the signal is highly dependent on the
higgsino mass. To further enhance the sensitivity of the analysis, we therefore subdivide the search region
into four \ptmiss bins: $150<\ptmiss\leq200\GeV$, $200<\ptmiss\leq300\GeV$, $300<\ptmiss\leq450\GeV$, and
$\ptmiss>450\GeV$.  The background estimation procedure described in Section~\ref{sec:backgroundEstimation}
is then applied separately in each of the four \ptmiss bins.

\section{Background estimation}
\label{sec:backgroundEstimation}

\subsection{Method}
\label{ssec:bkgest:overview}

\begin{figure}[tbp!]
 \includegraphics[width=\cmsFigPanel]{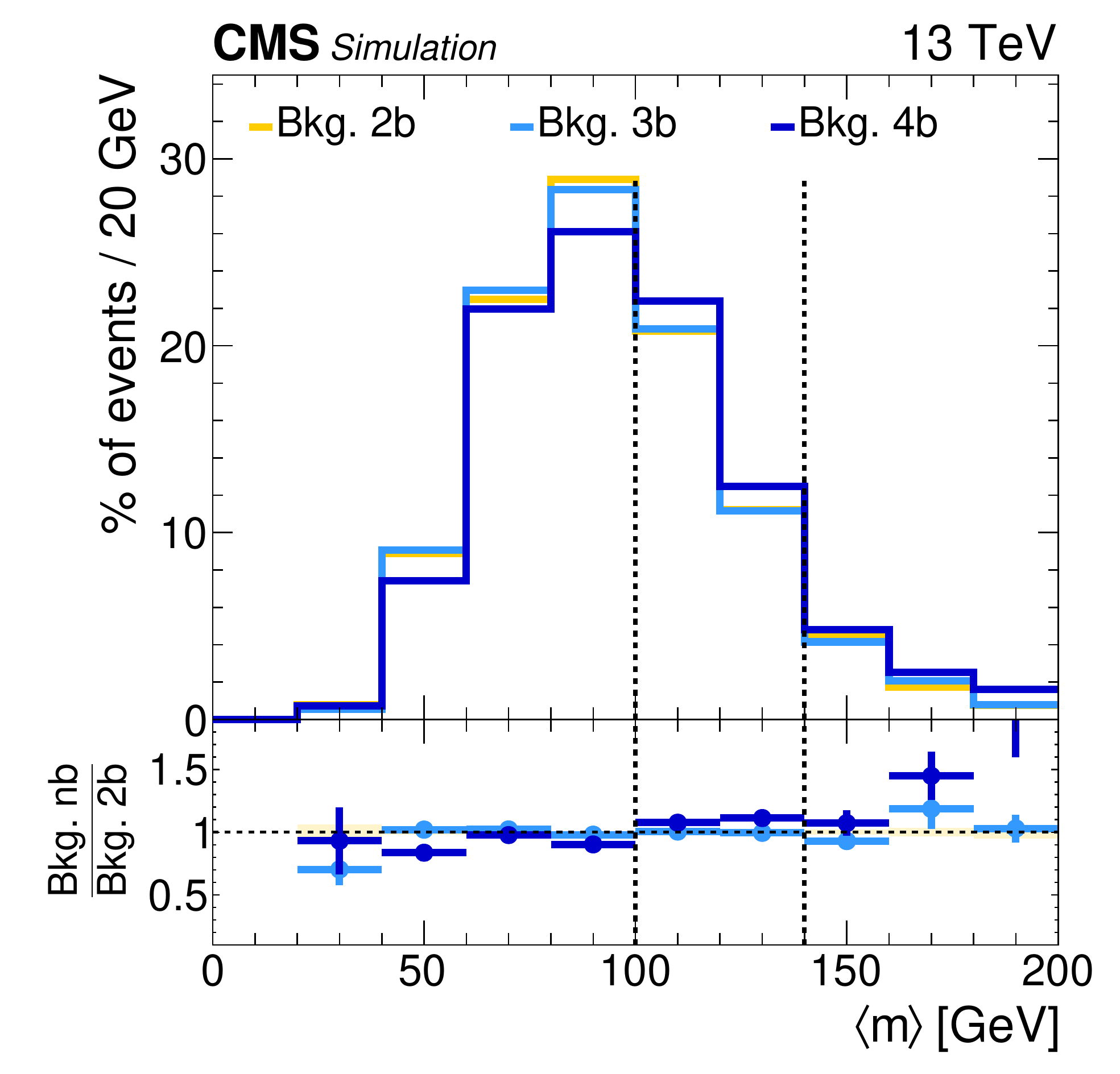}
\centering \caption{Distribution of \amjj after the baseline selection, showing the agreement between the
  \amjj shapes among the three \PQb tag categories. The comparison is based on simulation including all
  backgrounds except QCD multijet production, for which the simulation suffers from large statistical uncertainties. QCD multijet events account for less than 5\%
  of the total yield. The vertical dotted lines indicate the Higgs boson mass window.
\label{fig:bkgest:amjj_shapes}}
\end{figure}

The background estimation method is based on the observation that the \amjj distribution is approximately
uncorrelated with the number of \PQb tags. As shown in Fig.~\ref{fig:bkgest:amjj_shapes}, the \amjj shapes
for the three \PQb tag categories agree within the statistical uncertainty in the simulated samples. This
behavior can be understood by noting that the background in all three \PQb tag categories is dominated by events
containing only two \PQb quarks, with the additional b-tagged jets in the 3\PQb and  4\PQb  categories being
mistagged light-flavor or gluon jets. The background simulation indicates that only 20\%\,(37\%) of the events
in the 3\PQb ( 4\PQb ) selection have more than two \PQb quarks. As a result, the four jets used to construct \amjj
in the 3\PQb and  4\PQb  categories arise largely from the same fundamental processes as those with two b-tagged jets,
and thus the shape of the average mass distribution is independent of \nb for $\nb\geq2$.

Taking advantage of this observation, we estimate the background contribution to each signal bin with an ABCD
method~\cite{Khachatryan:2016uwr} that employs \amjj and the \PQb tag categories as the two ABCD variables
similarly to the 8\TeV analysis~\cite{Khachatryan:2014mma}. We define the Higgs boson mass window (HIG region)
as the events with \amjj within 100 to 140\GeV, and the Higgs boson mass sideband (SBD region) as the events
with {$0<\amjj<100$\GeV} or {$140<\amjj<200\GeV$}. The mass window is chosen to optimize the
signal sensitivity, taking into account the background distribution and the asymmetry in the Higgs boson mass
resolution. The 3\PQb and  4\PQb  SBD regions, together with the shape of the \amjj distribution in the 2\PQb category,
are then used to determine the background in the signal-enriched 3\PQb and  4\PQb  HIG regions independently for each
\ptmiss bin as follows:
\begin{linenomath*}
\begin{equation}
\label{eq:bkgest:abcd_ideal}
  \mu^\text{bkg}_{\mathrm{HIG,3\PQb}} = R \,\mu^\text{bkg}_{\mathrm{SBD,3\PQb}}
  \quad \text{and} \quad
  \mu^\text{bkg}_{\mathrm{HIG, 4\PQb }} = R \, \mu^\text{bkg}_{\mathrm{SBD, 4\PQb }}.
\end{equation}
\end{linenomath*}

Here, $\mu_{\mathrm{SBD,}n\PQb}^\text{bkg}$ and $\mu_{\mathrm{HIG,}n\PQb}^\text{bkg}$ are the background rates
for each \PQb tag category ($n=2,3,4$) in the SBD and HIG search regions, respectively, and $R$ is the ratio of
the background rate in the HIG region to that in the SBD region. In the limit that the \PQb tag category and
\amjj are uncorrelated, $R$ is the same for the three \PQb tag categories:
\begin{linenomath*}
\begin{equation}
R  \equiv
\left( \frac{  \mu^\text{bkg}_\mathrm{HIG}  }{  \mu^\text{bkg}_\mathrm{SBD}  }\right)_\mathrm{2\PQb}  =
\left( \frac{  \mu^\text{bkg}_\mathrm{HIG}  }{  \mu^\text{bkg}_\mathrm{SBD}  }\right)_\mathrm{3\PQb}  =
\left( \frac{  \mu^\text{bkg}_\mathrm{HIG}  }{  \mu^\text{bkg}_\mathrm{SBD}  }\right)_\mathrm{ 4\PQb }.
\end{equation}
\end{linenomath*}

\begin{figure*}[tbp!]
 \includegraphics[width=\cmsFigKappa]{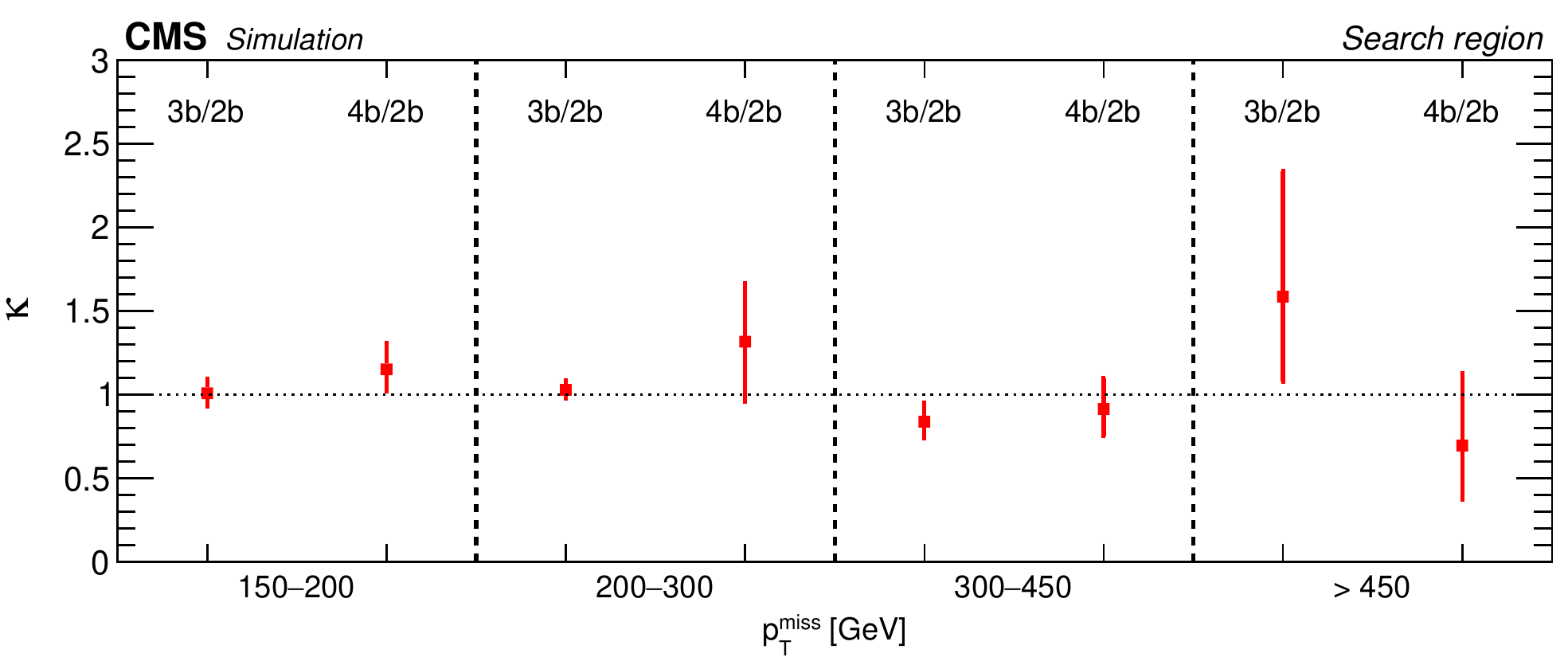}
 \centering
 \caption{Values of the double ratios $\kappa_{3\text{b}}$ and $\kappa_{4\text{b}}$ obtained from the background
   simulation for each of the \ptmiss bins. The error bars correspond to the statistical uncertainty of the background simulation.}
 \label{fig:bkgest:search_mckappas}
\end{figure*}

The \textit{closure} of the background estimation method, that is, the ability of Eq.~\eqref{eq:bkgest:abcd_ideal}
to predict the background rates in the signal regions, is quantified with the double ratios
\ifthenelse{\boolean{cms@external}}{
\begin{linenomath*}
\begin{equation} \label{eq:bkgext:kappa}
\begin{aligned}
\kappa_{3\PQb} &= \left( \frac{\mu^\text{bkg}_\mathrm{HIG}}{\mu^\text{bkg}_\mathrm{SBD}}\right)_{3\PQb}\Big/\left( \frac{\mu^\text{bkg}_\mathrm{HIG}}{\mu^\text{bkg}_\mathrm{SBD}}\right)_\mathrm{2\PQb}\\
\text{and}\\
\kappa_{4\PQb} &= \left( \frac{\mu^\text{bkg}_\mathrm{HIG}}{\mu^\text{bkg}_\mathrm{SBD}}\right)_{4\PQb}\Big/\left( \frac{\mu^\text{bkg}_\mathrm{HIG}}{\mu^\text{bkg}_\mathrm{SBD}}\right)_\mathrm{2\PQb}.\\
\end{aligned}
\end{equation}
\end{linenomath*}
}{
\begin{linenomath*}
\begin{equation} \label{eq:bkgext:kappa}
\kappa_{3\PQb} = \left( \frac{\mu^\text{bkg}_\mathrm{HIG}}{\mu^\text{bkg}_\mathrm{SBD}}\right)_{3\PQb}\Big/\left( \frac{\mu^\text{bkg}_\mathrm{HIG}}{\mu^\text{bkg}_\mathrm{SBD}}\right)_\mathrm{2\PQb}
\quad\text{and} \quad
\kappa_{4\PQb} = \left( \frac{\mu^\text{bkg}_\mathrm{HIG}}{\mu^\text{bkg}_\mathrm{SBD}}\right)_{4\PQb}\Big/\left( \frac{\mu^\text{bkg}_\mathrm{HIG}}{\mu^\text{bkg}_\mathrm{SBD}}\right)_\mathrm{2\PQb}.
\end{equation}
\end{linenomath*}
}
These \kapa factors measure the impact of any residual correlation between the \PQb tag category and \amjj.
Figure~\ref{fig:bkgest:search_mckappas} shows that the \kapa factors in simulation are consistent with unity
for both the 3\PQb and  4\PQb  regions across the full \ptmiss range, demonstrating the validity of the fundamental
assumption of the ABCD method. In Section~\ref{sec:syst}, we study the closure of the method in data control
samples and estimate the associated systematic uncertainties in the background prediction.

\subsection{Implementation}
\label{ssec:bkgdEstimationProcedure}

The method outlined in Section~\ref{ssec:bkgest:overview} is implemented with a likelihood function that
incorporates the statistical and systematic uncertainties associated with the background prediction and
the signal model, and also accounts for signal contamination in all control regions.

The terms in the likelihood function corresponding to the observed yields in all analysis regions, reflecting
the parameterization of the ABCD method and the signal contributions to each bin, can be written as the
following product of Poisson probability density functions (pdfs):
\ifthenelse{\boolean{cms@external}}{
\begin{linenomath*}
\begin{multline}
 \mathcal{L}_\mathrm{ABCD} =\\
  \prod_{m=1}^{4} \prod_{n=2}^{4}
 \text{Poisson}(N^\text{data}_{\mathrm{SBD,}n\PQb,m}|
 \mu_{\mathrm{SBD,}n\PQb,m}^{\text{bkg}} +r\,\mu_{\mathrm{SBD,}n\PQb,m}^{\text{sig}}) \\
\times \prod_{m=1}^{4} \prod_{n=2}^{4}
 \text{Poisson}(N^\text{data}_{\mathrm{HIG,}n\PQb,m}|
 R \,\mu_{\mathrm{SBD,}n\PQb,m}^{\text{bkg}} +r\,\mu_{\mathrm{HIG,}n\PQb,m}^{\text{sig}}). \label{eq:bkgest:likelihood}
\end{multline}
\end{linenomath*}
}{
\begin{linenomath*}
\begin{multline}
 \mathcal{L}_\mathrm{ABCD} = \prod_{m=1}^{4} \prod_{n=2}^{4}
 \text{Poisson}(N^\text{data}_{\mathrm{SBD,}n\PQb,m}|
 \mu_{\mathrm{SBD,}n\PQb,m}^{\text{bkg}} +r\,\mu_{\mathrm{SBD,}n\PQb,m}^{\text{sig}}) \\
\times \prod_{m=1}^{4} \prod_{n=2}^{4}
 \text{Poisson}(N^\text{data}_{\mathrm{HIG,}n\PQb,m}|
 R \,\mu_{\mathrm{SBD,}n\PQb,m}^{\text{bkg}} +r\,\mu_{\mathrm{HIG,}n\PQb,m}^{\text{sig}}). \label{eq:bkgest:likelihood}
\end{multline}
\end{linenomath*}
}
Here, the index $m$ runs over the four \ptmiss bins, the index $n$ runs over the three \PQb tag categories,
$N^\text{data}$ are the observed data yields, $\mu^{\text{sig}}$ are the expected signal rates, and $r$ is the
parameter quantifying the signal strength relative to the expected yield across all analysis bins. The four
main floating parameters describing the fitted background for each \ptmiss bin $m$ are the three sideband
background rates $\mu_{\mathrm{SBD,}n\PQb,m}^{\text{bkg}}$ and the ratio $R$.

The full likelihood function is given by the product of $\mathcal{L}_\mathrm{ABCD}$, Poisson pdfs constraining
the signal shape and its statistical uncertainty in each bin, and log-normal pdfs constraining nuisance
parameters that account for the systematic uncertainties in the closure and the signal efficiency. These
nuisance parameters were omitted from Eq.~\eqref{eq:bkgest:likelihood} for simplicity.

Following the approach in Ref.~\cite{Khachatryan:2016uwr}, the likelihood function is employed in two types of
fits: the \textit{predictive fit}, which allows us to more easily establish the agreement of the background
predictions and the observations in the background-only hypothesis, and the \textit{global fit}, which enables
us to estimate the signal strength.

The predictive fit is realized by removing the terms of the likelihood corresponding to the observed yields in
the signal regions, (HIG,~3\PQb) and (HIG,~4\PQb ), and fixing the signal strength $r$ to 0. As a result, we obtain a
system of equations with an equal number of unknowns and constraints. For each \ptmiss bin, the four main
floating parameters $\mu_{\mathrm{SBD},2\PQb}^\mathrm{bkg}$, $\mu_{\mathrm{SBD},3\PQb}^\mathrm{bkg}$, $\mu_{\mathrm{SBD}, 4\PQb }^\text{  bkg}$, and $R$ are determined by the four observations $N^\text{data}_{\mathrm{SBD,2\PQb}}$, $N^\text{data}_{\mathrm{    SBD},3\PQb}$, $N^\text{data}_{\mathrm{SBD}, 4\PQb }$, and $N^\text{data}_{\mathrm{HIG},2\PQb}$. Since the extra floating
parameters corresponding to the systematic uncertainties are constrained by their respective log-normal pdfs,
they do not contribute as additional degrees of freedom. The predictive fit thus converges to the standard
ABCD method, and the likelihood maximization machinery becomes just a convenient way to solve the system of
equations and to propagate the various uncertainties.

Conversely, the global fit includes the observations in the signal regions. Since in this case there are six
observations and four floating background parameters in each \ptmiss bin, there are enough constraints for the
signal strength $r$ to be determined in the fit. The global fit also properly accounts for the signal yields
in the control regions, using the signal shape across control and signal regions from the simulation.

\section{Systematic uncertainties in the background prediction}
\label{sec:syst}
\label{ssec:syst_intro}

The background estimation procedure described in Section~\ref{sec:backgroundEstimation} relies on the
approximate independence of the \amjj and \nb distributions. In
Sections~\ref{ssec:syst_ttbar}, \ref{ssec:syst_zjets}, and \ref{ssec:syst_qcd} we study this assumption for
individual background processes in data and simulation by defining dedicated control regions
for \ttbar, \zjets, and QCD multijet production. The overall level of closure in these control samples, better
than 13\% in all cases, is assigned as a systematic uncertainty for each of the main background sources
separately.  Additionally, these samples validate the closure in the simulation as a function of \ptmiss.

If the background estimation method is valid for each separate background contribution, then it would also be
valid for the full background composition as long as the relative abundance of each background component is
independent of \nb. In Section~\ref{ssec:syst_comp}, we use these data control samples to quantify the
validity of the simulation prediction that the background composition is independent of \nb in each \ptmiss
bin by examining the modeling of the \ptmiss and \nb distributions for each background source.

Finally, in Section~\ref{ssec:syst_summary}, we describe the prescription for assigning the total systematic
uncertainty in the background prediction binned in \ptmiss and \nb, taking into account both the findings from
the data control sample studies and the closure of the method in the simulation. The latter is the dominant
systematic uncertainty in this analysis.

\subsection{Single-lepton \texorpdfstring{\ttbar}{ttbar} control sample}
\label{ssec:syst_ttbar}

To test whether the background estimation method works for \ttbar events, we define a single-lepton control
sample, which, like the search region, is dominated by single-lepton \ttbar events and represents a similar
kinematic phase space. Because the lepton is a spectator object as far as the ABCD method is concerned---it is
neither involved in the construction of the \amjj variable, nor correlated with the presence of
additional \PQb tags---this control sample should accurately capture any potential mismodeling of
the \amjj-\nb correlation that may be present in the signal region. While this control region does not
specifically probe semileptonic \ttbar events involving a hadronically decaying $\tau$ lepton, the simulation
shows that their \amjj distribution in the signal region is the same as that of semileptonic \ttbar events involving
light leptons. This is expected because in most cases the $\tau$ lepton in these events is either out of
acceptance or reconstructed as the jet with the smallest \PQb-discriminator value and, as a result, it does not
enter the \amjj calculation.

\begin{figure}[tbp!]
  \centering
\includegraphics[width=\cmsFigPanel]{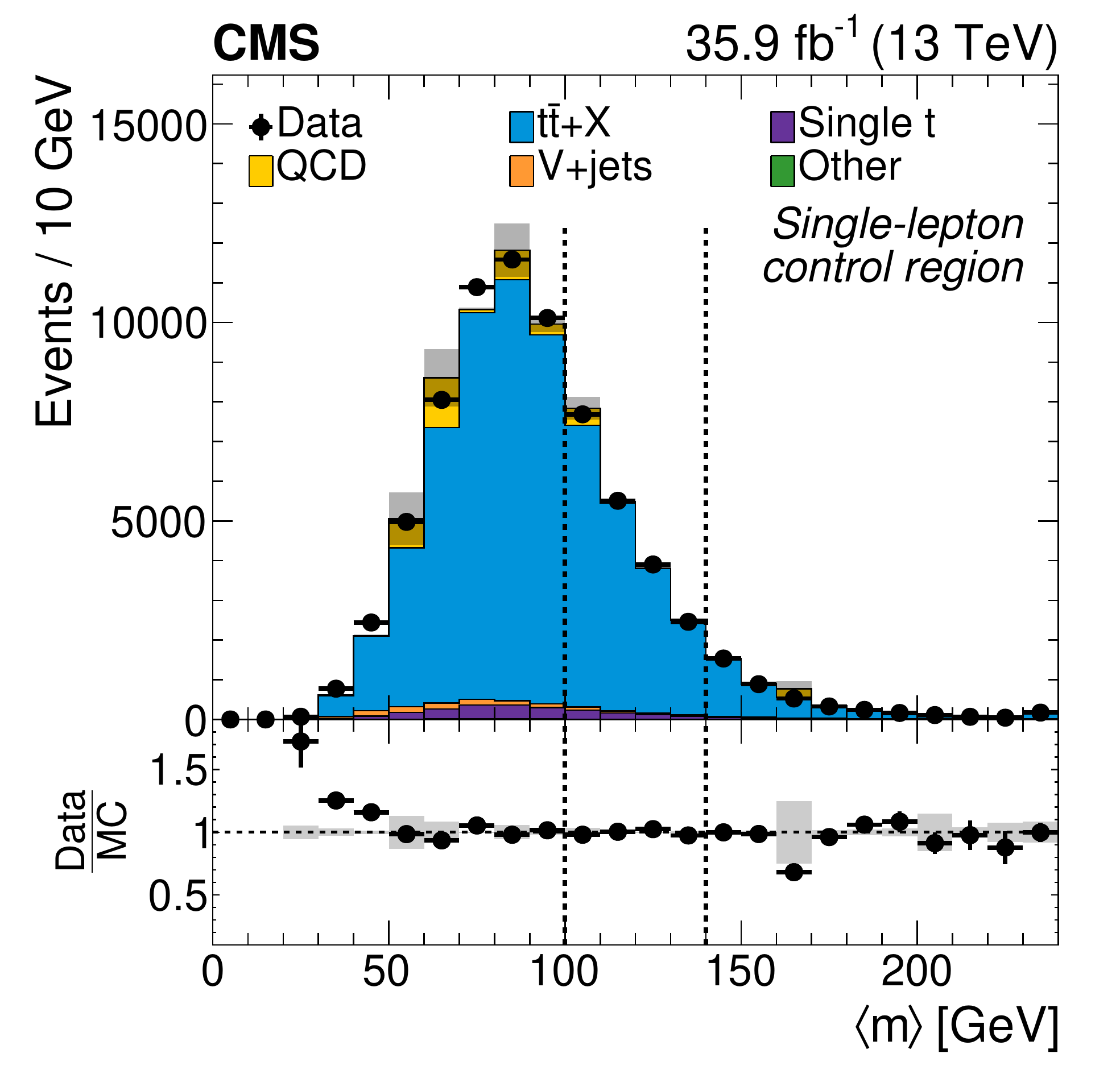}
\includegraphics[width=\cmsFigPanel]{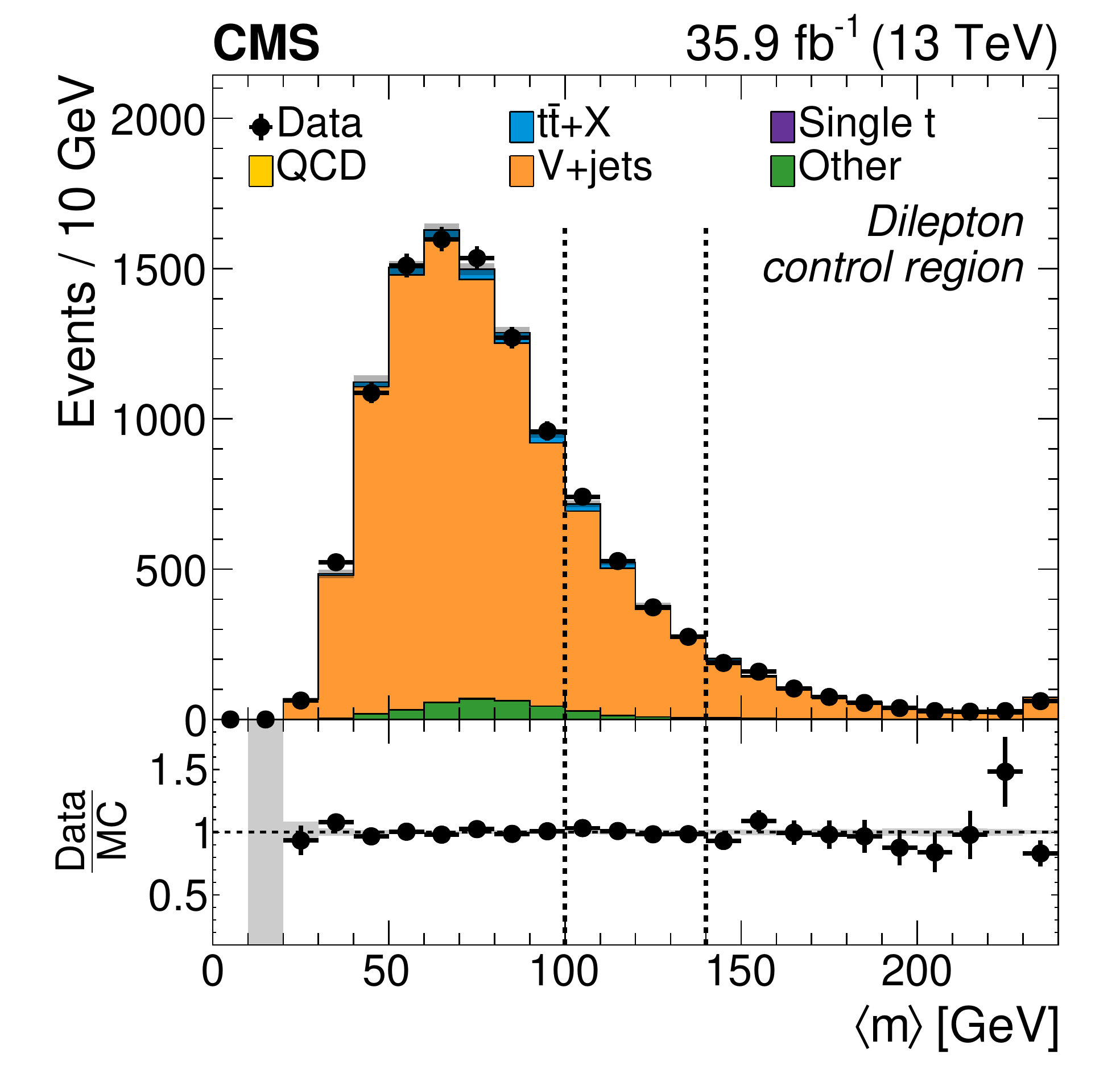}
  \caption{Comparison of the distributions of \amjj in data and simulation in the single-lepton control sample
    (\cmsLeft) and in the dilepton control sample (\cmsRight), where in both cases we have integrated
    over \ptmiss and \PQb tag categories. The overall yields in simulation have been normalized to those
    observed in data. The gray shading indicates the statistical uncertainty in the total simulated
    background. The last bin includes overflow.}
  \label{fig:syst:ttbar_datamc}
\end{figure}

For each of the four \ptmiss search bins, we construct a corresponding ABCD test in a single-lepton control
sample, defined by the same selection requirements except for removing the lepton and the isolated track
vetoes and instead requiring exactly one lepton with $\pt>30$\GeV (to reach trigger efficiency plateau) and
$\mt(\vec{p}^{\ell}_\mathrm{T},\ptvecmiss)<100$\GeV (to avoid poorly reconstructed events).
Given that the contamination from QCD multijet production in the
single-lepton region is small, the $\Delta\phi$ requirement is also removed to further improve the statistical
power of the control sample. Since the lepton provides a way to trigger on events with lower \ptmiss, we add
two additional \ptmiss bins, $\ptmiss<75$\GeV and {$75<\ptmiss\leq150$\GeV}, allowing us to study the
\ptmiss dependence of the closure in a wider range. In this control region, \ttbar production accounts for
over 90\% of the events, except for the two highest \ptmiss bins, where the total contribution of single top
quark and V$+$jets production can be as high as $\sim$25\%. Figure~\ref{fig:syst:ttbar_datamc} (\cmsLeft) shows
the comparison of the \amjj shapes between the data and the simulation.

\begin{figure*}[tbp!]
  \centering
  \includegraphics[width=\cmsFigKappa]{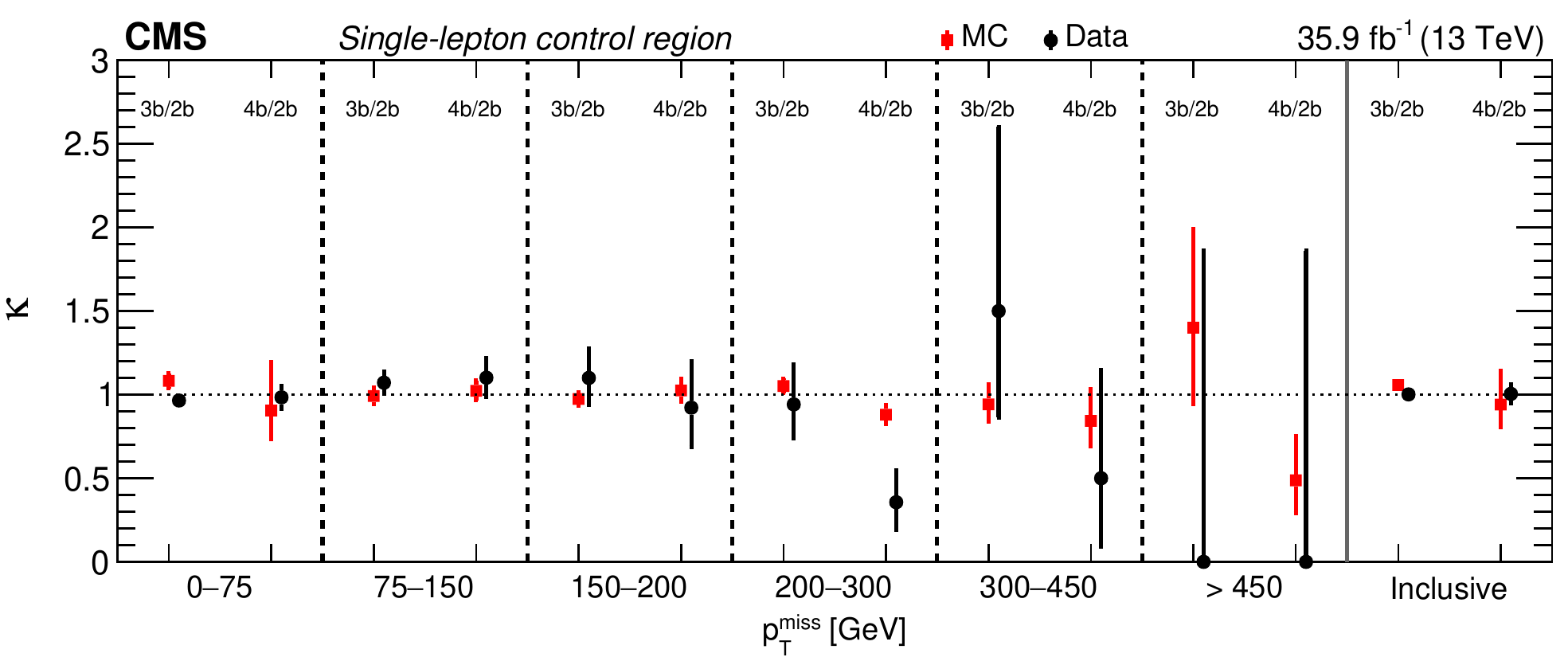}
  \caption{Comparison of the \kapa values found in the single-lepton control sample, for data and simulated
    events, for the 3\PQb/2\PQb and  4\PQb /2\PQb ABCD tests in each \ptmiss bin as well as after integrating over \ptmiss
    (labeled as ``Inclusive'').}
  \label{fig:syst:ttbar_datakappas}
\end{figure*}

As described in Section~\ref{sec:backgroundEstimation}, since the 3\PQb and  4\PQb  categories are dominated by events
with two true \PB\ hadrons and one or two additional mistagged jets, similar jet topologies contribute to all
b tag categories and thus the \amjj distributions of the reconstructed \PQb tag categories converge. We
validate this assertion in the single-lepton control sample by examining the value of the \kapa factors.
Figure~\ref{fig:syst:ttbar_datakappas} shows the overall closure of the method across bins of \ptmiss, both in
the simulation and in data. We observe agreement within the statistical uncertainties, with \kapa values being
consistent with unity across the full \ptmiss range for both data and simulation. This observation is also
confirmed with larger statistical precision by repeating the test in a more inclusive sample obtained by
removing the \drmax requirement.

An overall uncertainty in the validity of the method in \ttbar-like events is assigned based on the larger of
the nonclosure and the statistical uncertainty in the closure test in data performed after integrating over the full
\ptmiss range. The results, shown to the right of the solid line in Fig.~\ref{fig:syst:ttbar_datakappas},
correspond to an uncertainty of 3\% and 6\% in the 3\PQb and  4\PQb  bins, respectively.

\subsection{Dilepton \texorpdfstring{\zjets}{Z+jets} control sample}
\label{ssec:syst_zjets}

As shown in Table~\ref{tab:selection:cutflow}, the second-largest background is \zjets, with the \PZ boson
decaying via \znn. Similarly to the \ttbar case, we can validate the background estimation method for \zjets
events by constructing a closure test in a representative data control sample rich in \zll decays. However,
given the small branching fraction of \zll decays and the large \ttbar contamination associated with a high-\nb
selection, we validate the method by constructing ABCD tests at lower \PQb tag requirements, namely 1\PQb/0\PQb and
2\PQb/1\PQb.

The \zll control sample is constructed in a similar manner to the search region. Events with 4 or 5 jets are
selected, and the reconstruction of a pair of Higgs bosons proceeds as described in Section~\ref{sec:objects}.
We require two opposite-charge same-flavor signal leptons in the \PZ boson mass window, $80<\mll\leq100$\GeV,
with the \pt of the leading and subleading lepton required to be greater than 40 and 20\GeV, respectively. We
remove the lepton and isolated track vetoes and, since the dilepton selection makes the contamination from QCD multijet
events negligible, we remove the $\Delta\phi$ requirement.  Since we do not expect genuine \ptmiss in \zll
events, we additionally require {$\ptmiss<50$\GeV}, which reduces the contamination from other processes
from 20\% to 10\%.

We divide the sample in bins of \ptll, ensuring kinematic correspondence with the \znn decays present in the
various \ptmiss bins employed in the search region.  Similarly to the single-lepton sample, the presence of
leptons allows us to extend the closure test to lower values of \ptll. Figure~\ref{fig:syst:ttbar_datamc}
(\cmsRight) shows both the high purity of the sample and the excellent data-to-simulation agreement in the \amjj
shape.

\begin{figure*}[tbp!]
  \centering
  \includegraphics[width=\cmsFigKappa]{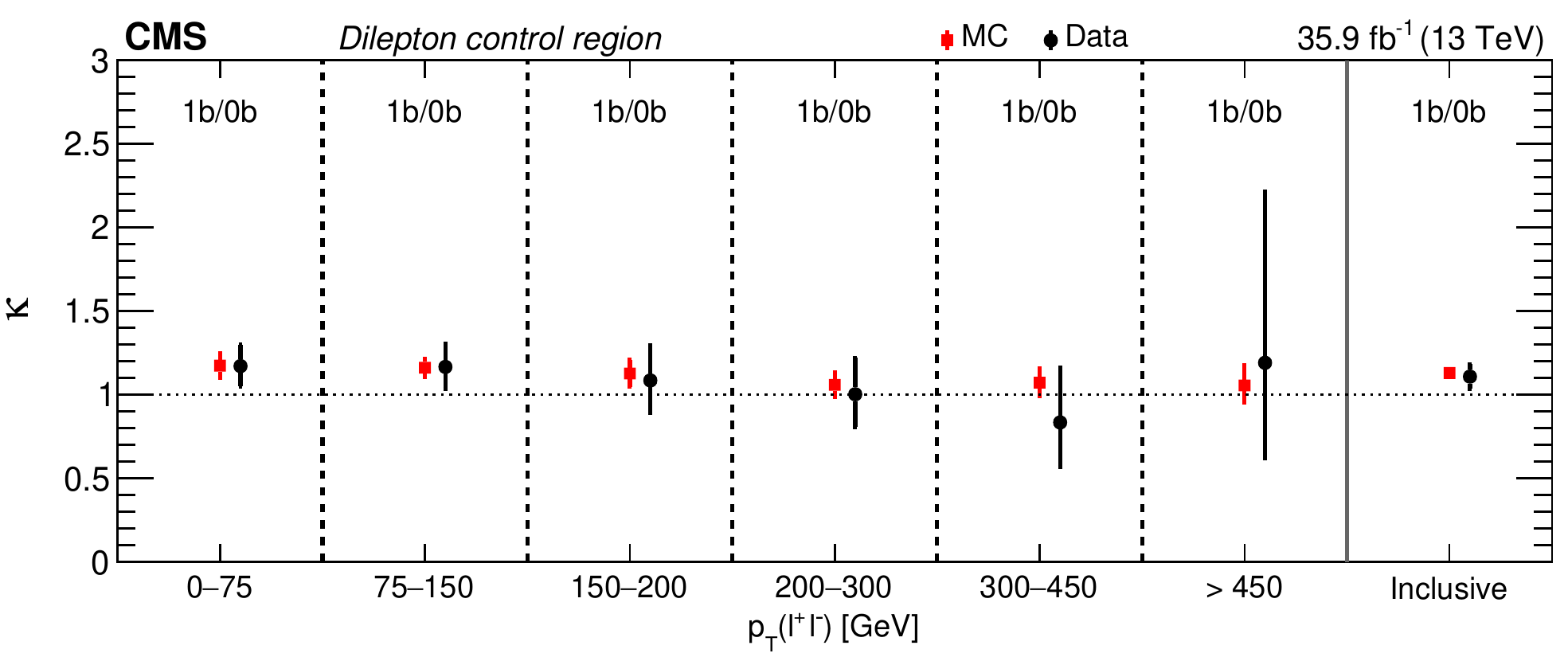}
  \caption{Comparison of the \kapa values found in the dilepton control sample, data and simulation, for the
    1\PQb/0\PQb ABCD tests in bins of \ptll as well as after integrating over \ptll (labeled as ``Inclusive'').}
  \label{fig:syst:zll_datakappas2b}
\end{figure*}

The validity of the extrapolation of the method to a sample consisting of lower \PQb tag multiplicities is
supported by the observation that all jets in \zjets events come from ISR, and thus their kinematic properties
are largely independent of the flavor content of the event. This expectation is confirmed in data by examining
the overall closure of the method in bins of \ptll for the 1\PQb/0\PQb and 2\PQb/1\PQb ABCD tests. The
1\PQb/0\PQb test, which has greater statistical power compared to the 2\PQb/1\PQb test and thus allows a
better examination of any potential trends as a function of \ptll, is shown in
Fig.~\ref{fig:syst:zll_datakappas2b} for illustration.

Since we do not observe that the closure of the method has any dependence on \ptll, we proceed to combine all
the \ptll bins into one bin and repeat the closure test with improved statistical precision. In the
1\PQb/0\PQb ABCD test, we observe a statistically significant nonclosure of 11\%, which may be due to higher
order effects beyond the precision of this search. The 2\PQb/1\PQb ABCD test shows good closure but with a
higher statistical uncertainty of 19\%. We assign the larger uncertainty of 19\% as the systematic uncertainty
in the closure of the background estimate method for \zjets events. The robustness of this result is further
corroborated by similar checks in a more inclusive selection without the \drmax requirement.

\subsection{Low \texorpdfstring{\dphi}{Dphi} QCD multijet control sample}
\label{ssec:syst_qcd}

Finally, to examine the validity of the ABCD method for QCD multijet events, we define a control region
enriched with such events by inverting the \dphi requirement. The high \PQb tag multiplicity region of this
control sample has a limited event yield and high \ttbar contamination. To overcome these limitations, we
exploit the fact that QCD multijet events, like \zjets events, have similar kinematic properties regardless of
their flavor content. We thus check the \amjj-\nb independence in lower \PQb tag multiplicity regions by
constructing the 1\PQb/0\PQb and 2\PQb/1\PQb ABCD tests.  We observe good agreement between the data and the
simulation for all \ptmiss bins. The maximum measured deviation of \kapa from unity in the inclusive bins
equals 13\%, which we assign as the systematic uncertainty in the closure of the background estimation method
for QCD multijet events.

\subsection{Impact of the background composition}
\label{ssec:syst_comp}

Having evaluated the closure of the method for each individual background, we proceed to study the impact of
mismodeling the relative abundance of the different background sources.

Since the \amjj shape varies among background types, as shown for \ttbar and \zjets in
Fig.~\ref{fig:syst:ttbar_datamc}, significant differences in the process admixture in the 2\PQb category with
respect to the 3\PQb or  4\PQb  category will result in \amjj-\nb correlation and lead to the nonclosure of the
method. From simulation, the background composition is expected to be independent of the \PQb tag category. The
validity of this prediction relies on the ability of the simulation to model the shape of the \PQb tag category
and \ptmiss distributions equally well for each background contribution.

From comparisons in the respective control samples, we indeed observe that the \nb distribution for each of
\ttbar, \zjets, and QCD multijet production is similarly well modeled by the simulation. The \ptmiss
distribution in simulation is found to overestimate the data for large values of \ptmiss for \ttbar and \zjets
events, while the opposite is observed for QCD multijet events. To provide an estimate of the impact of
mismodeling the background composition, we reweight the simulation based on the data-to-simulation comparisons
and then calculate the \kapa factors with the reweighted simulation, assigning 100\% of the shift with respect
to the nominal values as the uncertainty in the modeling of the background composition. The resulting
uncertainty is found to be at most 4\%.

\subsection{Total systematic uncertainty determination}
\label{ssec:syst_summary}

Based on the data control sample studies described in Sections~\ref{ssec:syst_ttbar}--\ref{ssec:syst_comp}, we assign a set of systematic uncertainties in the
background prediction for each search bin as follows:
\begin{enumerate}
  \item The closure uncertainty for each background process obtained in data control regions is propagated to
    the background predictions by varying the closure of the particular background in simulation in bins of
    \ptmiss and \nb. The resulting shifts on the predictions, ranging from 1\% to 10\% increasing with
    \ptmiss, are assigned as systematic uncertainties with a 100\% bin-to-bin correlation.
  \item The level of nonclosure due to the relative abundance of each background component as a function of
    \nb is estimated by comparing the change in \kapa in simulation before and after correcting the \nb and
    \ptmiss distributions of each background source according to measurements in the data control samples. Its
    overall impact is 1--4\% and it is taken as 100\% correlated across the different analysis bins.
  \item The closure studies in the data control samples with more inclusive selections show no evidence of
    \ptmiss dependence, but have insufficient statistical power at high \ptmiss using the default
    selection. Given this limitation and the extensive validation of the simulation in all control samples, we
    assign the larger of the statistical uncertainty and the nonclosure for each bin in the simulation as the
    systematic uncertainty in the background prediction as a function of \ptmiss and \nb. As seen in
    Fig.~\ref{fig:bkgest:search_mckappas}, this uncertainty ranges from 8--15\% in the lowest \ptmiss bin to
    59--75\% in the highest \ptmiss bin, and is assumed to be uncorrelated among bins.
\end{enumerate}

Each of the listed uncertainties is incorporated in the background fit as a log-normal constraint in the
likelihood function as described in Section~\ref{ssec:bkgdEstimationProcedure}, taking into account the stated
correlations. Because of the robustness of the background prediction method, evidenced by the high-statistics
data control region studies integrated in \ptmiss, the final uncertainty is dominated by the statistical
precision of the simulation in evaluating the closure as a function of \ptmiss, described in the third item.

\section{Results and interpretation}
\label{sec:results}
The observed event yields in data and the total predicted SM background are listed in
Table~\ref{tab:results:yields}, along with the expected yields for three higgsino mass scenarios.  Two
background estimates are given: the predictive fit, which does not use the data in the signal regions and
ignores signal contamination in the other regions, and the global fit with $r=0$, which incorporates the
observations in the (HIG,~3\PQb) and (HIG,~4\PQb ) regions, as described in
Section~\ref{ssec:bkgdEstimationProcedure}. Since for $\ptmiss>450$\GeV we observe no events in the (SBD,~4\PQb )
region, the parameter $\mu^\text{bkg}_{4\PQb,\mathrm{SBD}}$ is fitted to be zero, pushing against its physical
boundary and leading to the underestimation of the associated uncertainty. We account for this by including an
additional contribution that makes the uncertainty on $\mu^\text{bkg}_{4\PQb,\mathrm{SBD}}$ for
$\ptmiss>450$\GeV consistent with having observed one event. The event yields observed in data are consistent
with the background predictions for all the analysis bins  and no pattern of deviations is evident.

\begin{table*}[tbp]\centering
\topcaption{Event yields for all control regions---(HIG,~2\PQb), (SBD,~2\PQb), (SBD,~3\PQb), and (SBD,~4\PQb )---and
the two signal regions---(HIG,~3\PQb) and (HIG,~4\PQb )---in each of the four \ptmiss bins. The second column shows the
background yields predicted by the global fit which uses the observed yields in all control and signal regions, under the
background-only hypothesis ($r=0$). The third column gives the predicted SM background rates in the signal
regions obtained via the predictive fit which only takes as input the observed event yields in the control
regions. The expected signal yields for three signal benchmark points denoted as TChiHH(\mLSP,\mGold), with \mLSP and \mGold in units of \GeV, are also shown for reference.}
\renewcommand{\arraystretch}{1.2}
\label{tab:results:yields}
\begin{scotch}{ l  ccr  rrr}
Search & Global &        Predictive     & \multicolumn{1}{c}{Observed} & \multicolumn{1}{c}{TChiHH} & \multicolumn{1}{c}{TChiHH} & \multicolumn{1}{c}{TChiHH}        \\
region & fit & fit           & \multicolumn{1}{c}{yields} & \multicolumn{1}{c}{(225,1)} & \multicolumn{1}{c}{(400,1)} & \multicolumn{1}{c}{(700,1)}        \\ \hline

\multicolumn{7}{c}{$150 < \ptmiss \leq 200$\GeV}                                                      \\ \hline
SBD, 2\PQb & $1560.1^{+39.7}_{-38.5}$ &    \NA                  & 1559 & 5.8  & 1.3  & 0.0 \\
HIG, 2\PQb & $656.2^{+25.2}_{-24.6}$  &    \NA                  & 658  & 11.5 & 2.4  & 0.1 \\[1.2ex]
SBD, 3\PQb & $140.3^{+10.8}_{-10.3}$  &    \NA                  & 145  & 5.1  & 0.9  & 0.0 \\
HIG, 3\PQb & $57.7^{+5.5}_{-5.2}$     & $61.2^{+8.4}_{-7.7}$ & 53   & 10.9 & 2.5  & 0.1 \\[1.2ex]
SBD,  4\PQb  & $48.1^{+6.4}_{-5.8}$     &    \NA                  & 45   & 5.8  & 1.1  & 0.0 \\
HIG,  4\PQb  & $21.9^{+3.5}_{-3.2}$     & $19.0^{+4.6}_{-3.9}$ & 25   & 21.8 & 5.4  & 0.2 \\
  \hline

\multicolumn{7}{c}{$200 < \ptmiss \leq 300$\GeV}                                     \\ \hline
SBD, 2\PQb & $588.0^{+24.2}_{-23.5}$  &     \NA                 & 585  & 3.0  & 3.2  & 0.1 \\
HIG, 2\PQb & $333.1^{+17.9}_{-17.6}$  &     \NA                 & 336  & 6.0  & 6.6  & 0.3 \\[1.2ex]
SBD, 3\PQb & $55.3^{+6.5}_{-5.9}$     &     \NA                 & 61   & 2.2  & 2.6  & 0.1 \\
HIG, 3\PQb & $30.6^{+3.9}_{-3.6}$     & $35.1^{+5.9}_{-5.5}$ & 25   & 5.0  & 6.2  & 0.3 \\[1.2ex]
SBD,  4\PQb  & $15.6^{+3.8}_{-3.1}$     &     \NA                 & 13   & 2.4  & 3.3  & 0.1 \\
HIG,  4\PQb  & $11.4^{+3.0}_{-2.5}$     & $7.5^{+3.8}_{-2.7}$  & 14   & 14.3 & 15.7 & 0.8 \\
   \hline

\multicolumn{7}{c}{$300 < \ptmiss \leq 450$\GeV }                                    \\ \hline
SBD, 2\PQb & $72.4^{+8.7}_{-8.1}$     &     \NA                 & 74   & 0.0  & 1.9  & 0.2 \\
HIG, 2\PQb & $40.6^{+6.3}_{-6.0}$     &     \NA                 & 39   & 0.4  & 4.9  & 0.7 \\[1.2ex]
SBD, 3\PQb & $5.7^{+2.2}_{-1.8}$      &     \NA                 & 4    & 0.1  & 1.6  & 0.2 \\
HIG, 3\PQb & $3.3^{+1.4}_{-1.1}$      & $2.1^{+1.4}_{-1.0}$  & 5    & 0.9  & 4.6  & 0.5 \\[1.2ex]
SBD,  4\PQb  & $1.9^{+1.4}_{-0.9}$      &      \NA                & 2    & 0.2  & 1.5  & 0.2 \\
HIG,  4\PQb  & $1.1^{+0.8}_{-0.5}$      & $1.1^{+1.0}_{-0.6}$  & 1    & 2.0  & 10.3 & 1.5 \\
 \hline

\multicolumn{7}{c}{$\ptmiss > 450$\GeV }                                             \\ \hline
SBD, 2\PQb & $5.4^{+2.5}_{-2.1}$      &     \NA                 & 5    & 0.0  & 0.1  & 0.2 \\
HIG, 2\PQb & $4.6^{+2.2}_{-1.9}$      &     \NA                 & 5    & 0.0  & 0.4  & 0.9 \\[1.2ex]
SBD, 3\PQb & $0.6^{+0.8}_{-0.4}$      &     \NA                 & 1    & 0.1  & 0.1  & 0.2 \\
HIG, 3\PQb & $0.4^{+0.6}_{-0.3}$      & $1.0^{+1.6}_{-1.0}$  & 0    & 0.0  & 0.4  & 0.7 \\[1.2ex]
SBD,  4\PQb  & $0.0^{+0.3}_{-0.0}$      &    \NA                  & 0    & 0.0  & 0.1  & 0.2 \\
HIG,  4\PQb  & $0.0^{+0.3}_{-0.0}$      & $0.0^{+1.2}_{-0.0}$  & 0    & 0.0  & 1.1  & 2.0 \\
\end{scotch}
\end{table*}

\begin{figure*}[tbp!]
  \centering
  \includegraphics[width=\cmsFigPanel]{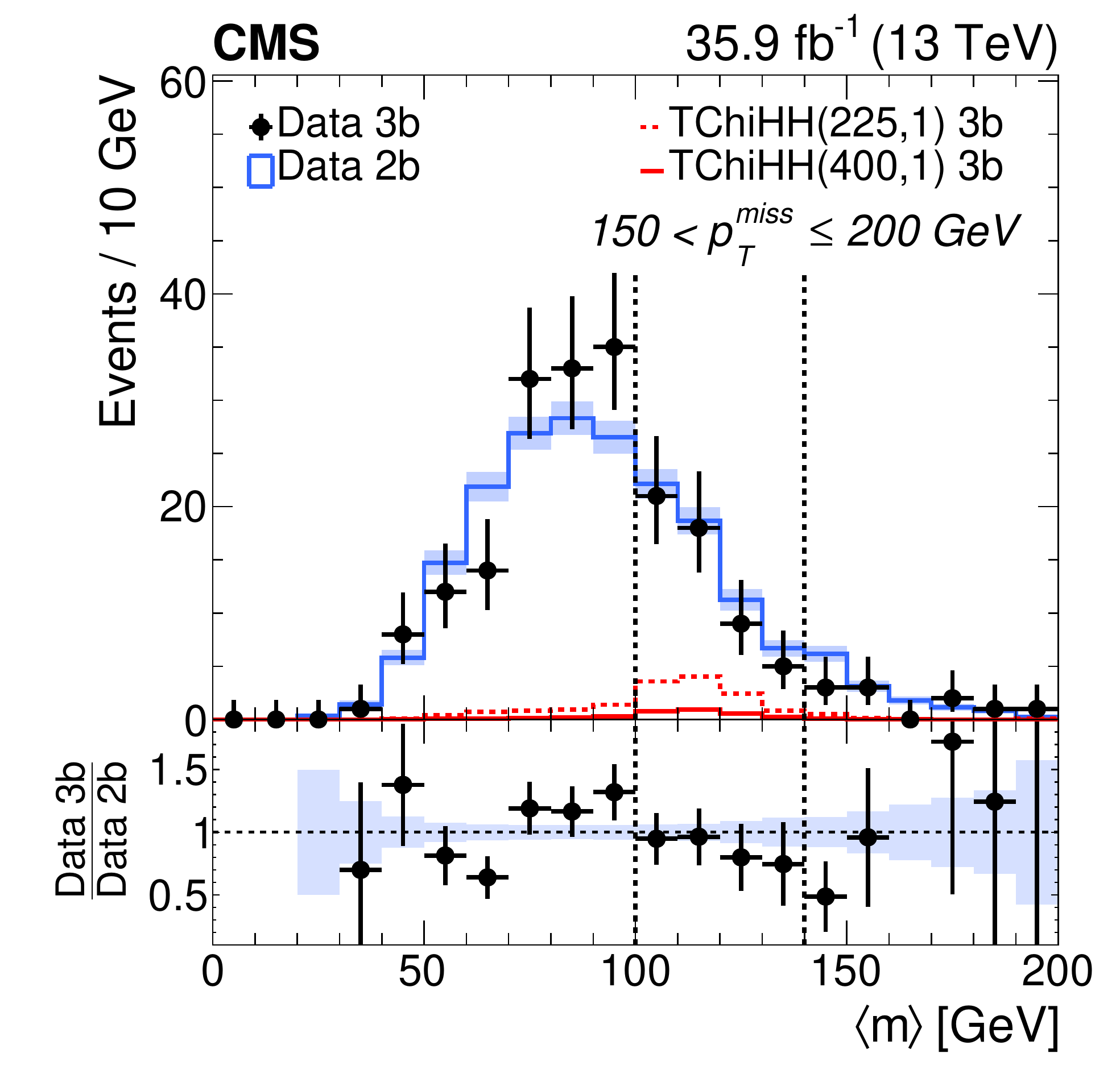}
  \includegraphics[width=\cmsFigPanel]{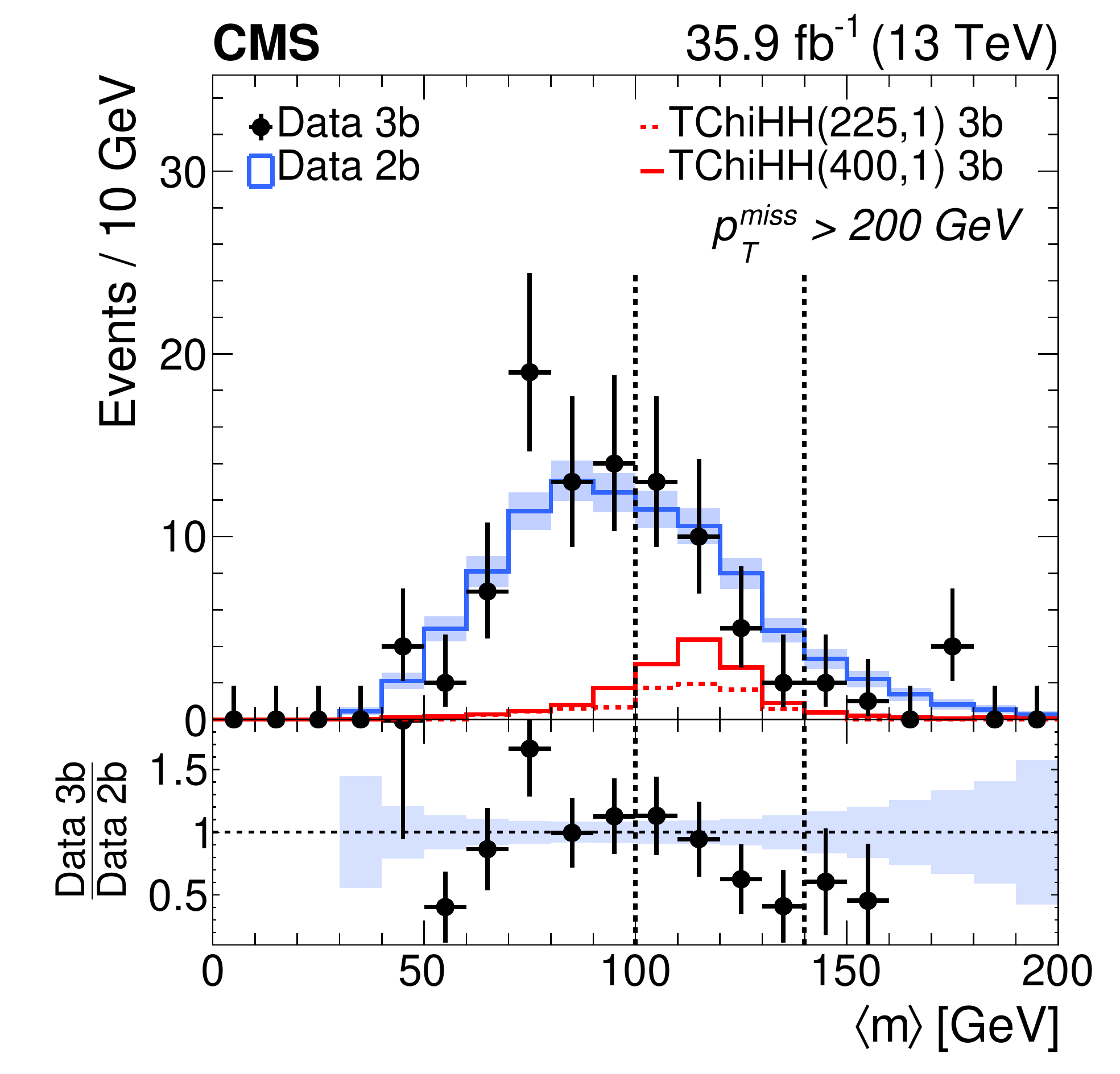}
  \includegraphics[width=\cmsFigPanel]{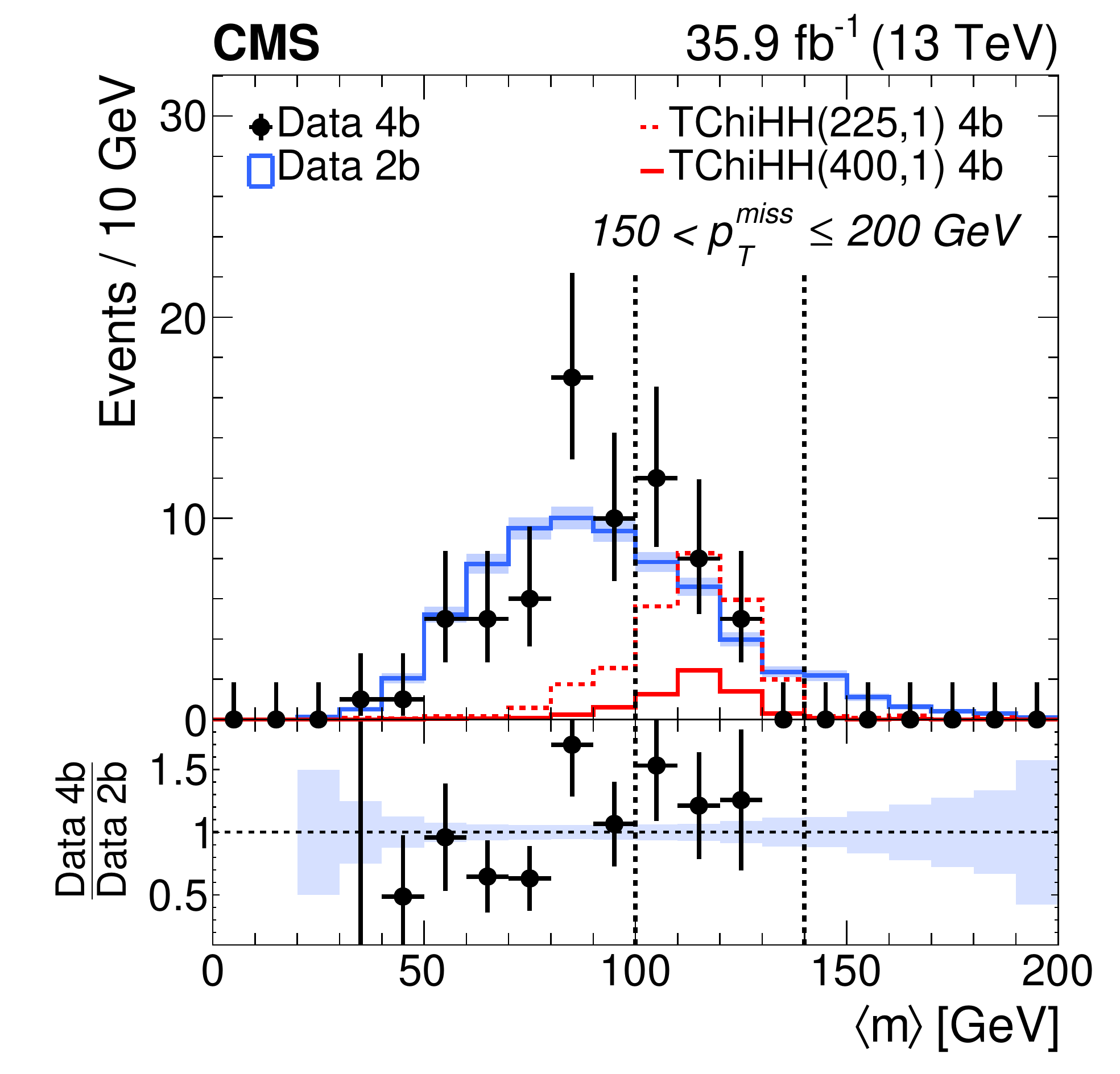}
  \includegraphics[width=\cmsFigPanel]{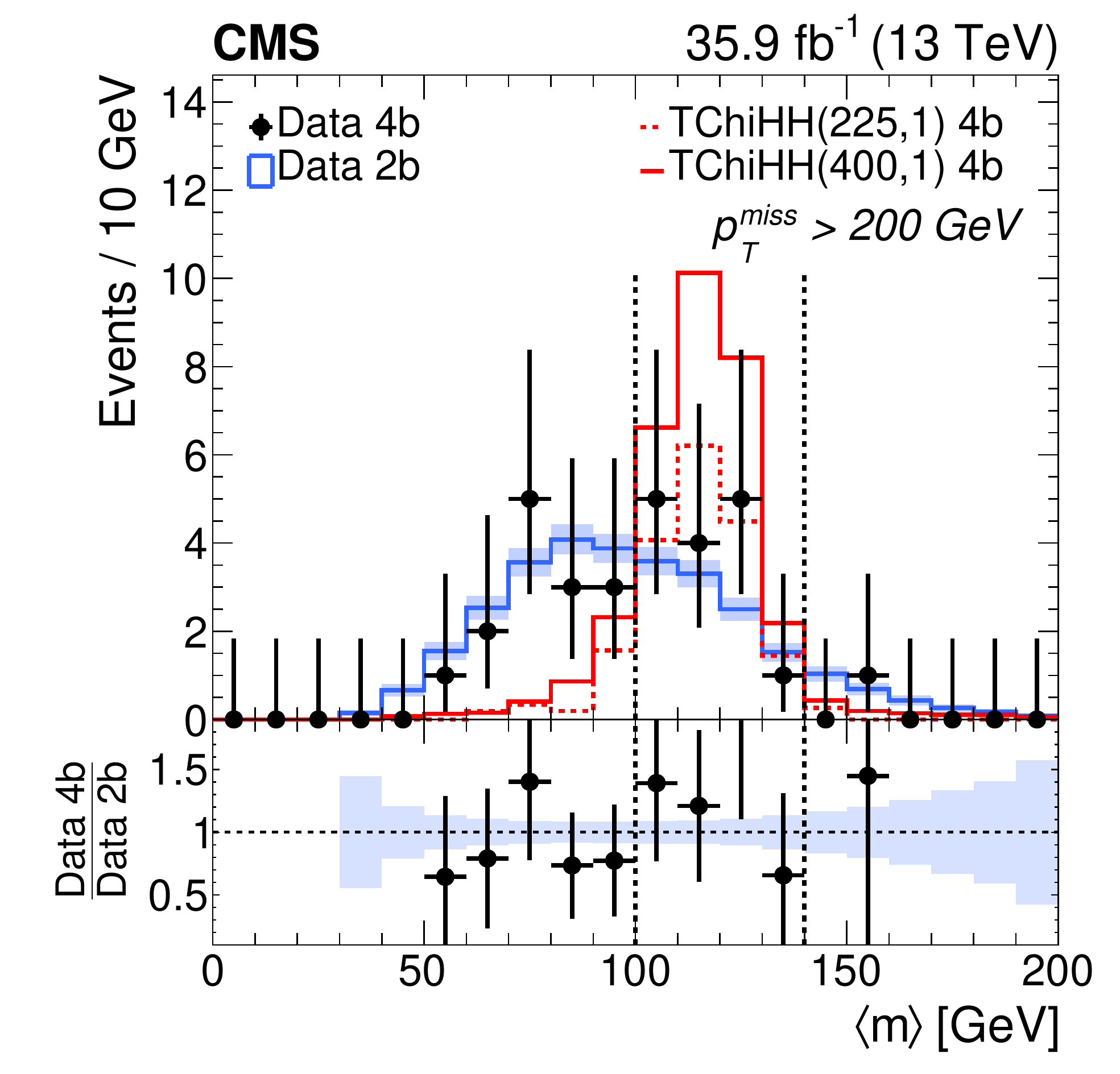}
  \caption{Distributions of \amjj in data and two signal benchmark models denoted as TChiHH(\mLSP,\mGold),
with \mLSP and \mGold in units of \GeV. The points with error bars show the data in the $3\PQb$ (top) and
$4\PQb$ bins (bottom) for $150<\ptmiss\leq200\GeV$ (left) and $\ptmiss>200\GeV$ (right).  The histograms show
the shapes of the \amjj distributions observed in the $2\PQb$ bins with overall event yields normalized to
those observed in the $3\PQb$ and $4\PQb$ samples. The shaded areas reflect the statistical uncertainty in
the \amjj distribution in the 2\PQb data. The vertical dashed lines denote the boundaries between the HIG and
the SBD regions. The ratio plots demonstrate that the shapes are in agreement.}
\label{fig:results:search_shapes_data}
\end{figure*}

Figure~\ref{fig:results:search_shapes_data} shows the distributions in data of \amjj in the $3\PQb$ and
$4\PQb$ bins for $150<\ptmiss\leq200\GeV$ and $\ptmiss>200\GeV$.  In each plot, the \amjj histogram
corresponding to the 2\PQb category is normalized to the integral of the overlaid high-\nb category distribution.
The shapes of the \amjj distributions are consistent and no significant excess is observed in either the 3\PQb or
the  4\PQb  HIG regions.

The absence of excess event yields in data is interpreted in the context of the higgsino simplified model
discussed in Section~\ref{sec:intro}. Table~\ref{tab:unc:sig} shows typical values for the systematic
uncertainties associated with the expected signal yields for three models with different higgsino masses. The
ranges correspond to the full variation of the uncertainties across all search bins. The uncertainty due to
the pileup modeling is given by the difference between the signal efficiencies evaluated in samples with the
mean number of reconstructed vertices found in the simulation and in the data, with the latter efficiencies obtained by
extrapolation. The evaluation of the pileup uncertainty for very low higgsino masses is limited by the
statistical power of the simulated samples.  The remaining uncertainties are determined by comparing the
nominal signal yield for each search region to the corresponding yield obtained after varying the scale factor
or correction under study within its uncertainty. In the case of the ISR uncertainty, the variation is based
on the full size of the ISR correction derived by comparing the transverse momentum of the jet system
balancing the $\cPZ$ boson in \zll events in data and in simulation. The largest uncertainties arise from the
jet energy corrections, jet energy resolution, pileup modeling, and the \ptmiss resolution in the fast
simulation. These uncertainties can be as large as 30\% for low higgsino masses, but their impact is smaller
for larger values of the higgsino mass. Uncertainties associated with the modeling of the \PQb tagging range from
1\% to 13\%. The uncertainties in the trigger efficiency range from 6\% in the lowest \ptmiss bin to $<$1\%
for $\ptmiss>300$\GeV. Uncertainties due to the modeling of ISR and the efficiency of the jet identification
filter are 1--2\%. Finally, the systematic uncertainty in the total integrated luminosity is
2.5\%~\cite{CMS-PAS-LUM-17-001}.

\begin{figure*}[tbp!]
\centering
\includegraphics[width=\cmsFigLimit]{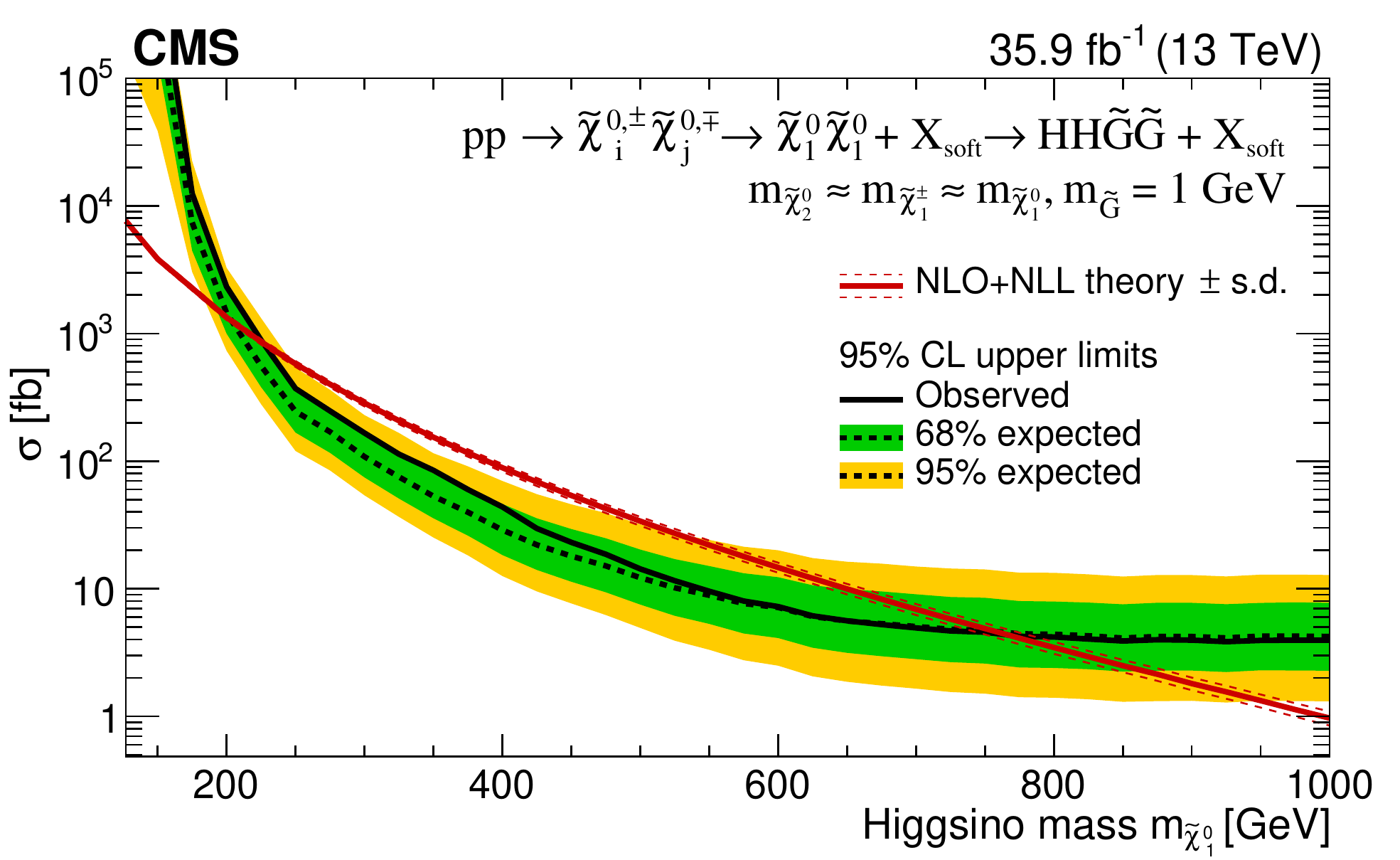}
\caption{Expected (dashed black line) and observed (solid black line) excluded cross sections at 95\% CL as a
function of the higgsino mass. The theoretical cross section for the TChiHH simplified model is shown as the
red solid line.}
\label{fig:results:limits}
\end{figure*}

The 95\% confidence level (\CL) upper limit on the production cross section for a pair of higgsinos in the
context of the TChiHH simplified model is estimated using the modified frequentist \cls
method~\cite{Junk:1999kv,Read:2002hq,CMS-NOTE-2011-005}, with a one-sided profile likelihood ratio test
statistic in its asymptotic approximation~\cite{Cowan:2010js}. Figure~\ref{fig:results:limits} shows the
expected and observed exclusion limits. The theoretical cross section at
NLO+NLL~\cite{Fuks:2012qx,Fuks:2013vua} as a function of higgsino mass is shown as a solid red line and the
corresponding uncertainty as a dotted red line. The upper limits on the cross section at 95\% CL for each mass
point are obtained from the global fit method, which takes into account the expected signal contribution in
all of the bins. Higgsinos with masses between 230 and 770\GeV are excluded.

The sensitivity at low higgsino mass is limited by the acceptance of the \ptmiss triggers employed in this
analysis. As a result, final states corresponding to other Higgs boson decays that can be triggered
independently of \ptmiss, such as $\PH\to\gamma\gamma$~\cite{Sirunyan:2017eie} or
$\PH\to\PW\PW$~\cite{Sirunyan:2017lae}, become more important in the low-mass region. For high higgsino mass,
most of the signal events contribute to the highest
\ptmiss bin, which has a negligible amount of background, so the sensitivity is mainly limited by the cross
section for higgsino pair production.

\begin{table*}[tb]\centering
\topcaption{Range of values for the systematic uncertainties in the signal efficiency and acceptance across the
(HIG,~3\PQb) and (HIG,~4\PQb ) bins for three signal benchmark points denoted as TChiHH(\mLSP,\mGold),
with \mLSP and \mGold in units of \GeV.  Uncertainties due to a particular source are treated as fully
correlated among bins, while uncertainties due to different sources are treated as uncorrelated.}
\label{tab:unc:sig}
\renewcommand{\arraystretch}{1.1}
\begin{scotch}{lccc}
\multirow{2}{*}{Source}&\multicolumn{3}{c}{Relative uncertainty [\%]}\\
& \multicolumn{1}{c}{TChiHH(225,1)} & \multicolumn{1}{c}{TChiHH(400,1)} & \multicolumn{1}{c}{TChiHH(700,1)}  \\
  \hline
Trigger efficiency             & 1--6   & 1--6  & 1--6   \\
b tagging efficiency           & 1--5   & 1--5  & 2--5   \\
Fast sim.~\PQb tagging efficiency & 2--10  & 4--8  & 3--13  \\
Fast sim.~\ptmiss resolution   & 14--27 & 1--6  & 1--7 \\
Jet energy corrections         & 8--32  & 4--22 & 2--12  \\
Jet energy resolution          & 3--23  & 1--18 & 1--11  \\
Initial-state radiation        & 1--2   & 1     & 1   \\
Jet identification             & 1      & 1     & 1      \\
Pileup                         & 1--31  & 1--6  & 1--5  \\
Integrated luminosity          & 3      & 3     & 3      \\
\end{scotch}
\end{table*}

\section{Summary}
\label{sec:summary}
A search for an excess of events is performed in proton-proton collisions in the channel with two Higgs bosons
and large missing transverse momentum (\ptmiss), with each of the Higgs bosons reconstructed in its
$\PH\to\bbbar$ decay. The data sample corresponds to an integrated luminosity of 35.9\fbinv at $\sqrt{s}
=13$\TeV. Because the signal has four $\PQb$ quarks, while the background is dominated by $\ttbar$ events
containing only two $\PQb$ quarks from the $\PQt$ quark decays, the analysis is binned in the number of
$\PQb$-tagged jets. In each event, the mass difference between the two Higgs boson candidates is required to
be small, and the average mass of the two candidates is used in conjunction with the number of observed
$\PQb$ tags to define signal and sideband regions. The observed event yields in these regions are used to
obtain estimates for the standard model background in the signal regions without input from simulated event
samples. The data are also binned in regions of $\ptmiss$ to enhance the sensitivity to the signal.

The observed event yields in the signal regions are consistent with the background predictions. These results
are interpreted in the context of a model in which each higgsino decays into a Higgs boson and a nearly
massless lightest supersymmetric particle (LSP), which is weakly interacting.  Such a scenario occurs in
gauge-mediated supersymmetry breaking models, in which the LSP is a goldstino. The cross section
calculation assumes that the higgsino sector is mass degenerate and sums over the cross sections for the pair
production of all relevant combinations of higgsinos, but all decays are assumed to be prompt. Higgsinos with
masses in the range 230 to 770\GeV are excluded at 95\% confidence level. These results constitute the most
stringent exclusion limits on this model to date.

\begin{acknowledgments}
We congratulate our colleagues in the CERN accelerator departments for the excellent performance of the LHC and thank the technical and administrative staffs at CERN and at other CMS institutes for their contributions to the success of the CMS effort. In addition, we gratefully acknowledge the computing centers and personnel of the Worldwide LHC Computing Grid for delivering so effectively the computing infrastructure essential to our analyses. Finally, we acknowledge the enduring support for the construction and operation of the LHC and the CMS detector provided by the following funding agencies: BMWFW and FWF (Austria); FNRS and FWO (Belgium); CNPq, CAPES, FAPERJ, and FAPESP (Brazil); MES (Bulgaria); CERN; CAS, MoST, and NSFC (China); COLCIENCIAS (Colombia); MSES and CSF (Croatia); RPF (Cyprus); SENESCYT (Ecuador); MoER, ERC IUT, and ERDF (Estonia); Academy of Finland, MEC, and HIP (Finland); CEA and CNRS/IN2P3 (France); BMBF, DFG, and HGF (Germany); GSRT (Greece); OTKA and NIH (Hungary); DAE and DST (India); IPM (Iran); SFI (Ireland); INFN (Italy); MSIP and NRF (Republic of Korea); LAS (Lithuania); MOE and UM (Malaysia); BUAP, CINVESTAV, CONACYT, LNS, SEP, and UASLP-FAI (Mexico); MBIE (New Zealand); PAEC (Pakistan); MSHE and NSC (Poland); FCT (Portugal); JINR (Dubna); MON, RosAtom, RAS, RFBR and RAEP (Russia); MESTD (Serbia); SEIDI, CPAN, PCTI and FEDER (Spain); Swiss Funding Agencies (Switzerland); MST (Taipei); ThEPCenter, IPST, STAR, and NSTDA (Thailand); TUBITAK and TAEK (Turkey); NASU and SFFR (Ukraine); STFC (United Kingdom); DOE and NSF (USA).

\hyphenation{Rachada-pisek} Individuals have received support from the Marie-Curie program and the European Research Council and Horizon 2020 Grant, contract No. 675440 (European Union); the Leventis Foundation; the A. P. Sloan Foundation; the Alexander von Humboldt Foundation; the Belgian Federal Science Policy Office; the Fonds pour la Formation \`a la Recherche dans l'Industrie et dans l'Agriculture (FRIA-Belgium); the Agentschap voor Innovatie door Wetenschap en Technologie (IWT-Belgium); the Ministry of Education, Youth and Sports (MEYS) of the Czech Republic; the Council of Science and Industrial Research, India; the HOMING PLUS program of the Foundation for Polish Science, cofinanced from European Union, Regional Development Fund, the Mobility Plus program of the Ministry of Science and Higher Education, the National Science Center (Poland), contracts Harmonia 2014/14/M/ST2/00428, Opus 2014/13/B/ST2/02543, 2014/15/B/ST2/03998, and 2015/19/B/ST2/02861, Sonata-bis 2012/07/E/ST2/01406; the National Priorities Research Program by Qatar National Research Fund; the Programa Clar\'in-COFUND del Principado de Asturias; the Thalis and Aristeia programs cofinanced by EU-ESF and the Greek NSRF; the Rachadapisek Sompot Fund for Postdoctoral Fellowship, Chulalongkorn University and the Chulalongkorn Academic into Its 2nd Century Project Advancement Project (Thailand); the Welch Foundation, contract C-1845; and the Weston Havens Foundation (USA).
\end{acknowledgments}

\bibliography{auto_generated}

\cleardoublepage \appendix\section{The CMS Collaboration \label{app:collab}}\begin{sloppypar}\hyphenpenalty=5000\widowpenalty=500\clubpenalty=5000\textbf{Yerevan Physics Institute,  Yerevan,  Armenia}\\*[0pt]
A.M.~Sirunyan, A.~Tumasyan
\vskip\cmsinstskip
\textbf{Institut f\"{u}r Hochenergiephysik,  Wien,  Austria}\\*[0pt]
W.~Adam, F.~Ambrogi, E.~Asilar, T.~Bergauer, J.~Brandstetter, E.~Brondolin, M.~Dragicevic, J.~Er\"{o}, M.~Flechl, M.~Friedl, R.~Fr\"{u}hwirth\cmsAuthorMark{1}, V.M.~Ghete, J.~Grossmann, J.~Hrubec, M.~Jeitler\cmsAuthorMark{1}, A.~K\"{o}nig, N.~Krammer, I.~Kr\"{a}tschmer, D.~Liko, T.~Madlener, I.~Mikulec, E.~Pree, D.~Rabady, N.~Rad, H.~Rohringer, J.~Schieck\cmsAuthorMark{1}, R.~Sch\"{o}fbeck, M.~Spanring, D.~Spitzbart, J.~Strauss, W.~Waltenberger, J.~Wittmann, C.-E.~Wulz\cmsAuthorMark{1}, M.~Zarucki
% ...
\vskip\cmsinstskip
% ...
1:~~Also at Vienna University of Technology, Vienna, Austria\\
% ...

\end{sloppypar}
\end{document}